\documentclass[pra,aps,a4paper,showpacs,superscriptaddress,floatfix,twocolumn]{revtex4-1}
\usepackage{amsmath}
\usepackage{amsfonts}
\usepackage{amssymb}
\usepackage{graphicx}
\usepackage{dcolumn}
\usepackage{bm}
\begin{document}

\def\Zz{\mathbb{Z}}
\def\T{\hat{T}}

\def\beq{\begin{equation}}
\def\eeq{\end{equation}}
\def\nn{\nonumber}
\def\om{\omega}
\def\a{\alpha}
\def\b{\beta}
\def\g{\gamma}
\def\th{\theta}
\def\eps{\epsilon}
\def\veps{\varepsilon}
\def\l{\lambda}
\def\p{\psi}

\def\s{\sigma}
\def\D{\Delta}
\def\d{\delta}

\def\S{\hat{S}}
\def\ph{\phantom}
\def\ra{\rightarrow}
\def\P{\hat{P}}
\def\M{\hat{M}}
\def\A{\bm{A}}
\def\v{\bm{v}}
\def\lo{\l^{(1)}}
\def\lz{\l^{(2)}}
\def\kt{{\tilde{k}}}
\def\sign{{\rm sign}}
\def\st{\mathfrak{s}}
\def\ka{\nu}
\def\asinh{{\rm asinh}}
\def\Vh{\frac{V}{2}}

\title{Integrability and weak diffraction in a two-particle Bose-Hubbard model}

\author{Daniel Braak}
\affiliation{Experimental Physics VI, Center for Electronic Correlations and Magnetism, University of Augsburg, 86135 Augsburg, Germany}

\author{J.~M.~Zhang}
\affiliation{Theoretical Physics III, Center for Electronic Correlations and Magnetism, University of Augsburg, 86135 Augsburg, Germany}
\affiliation{Max Planck Institute for the Physics of Complex Systems, N\"
othnitzer Str.~38, 01187 Dresden, Germany}

\author{Marcus Kollar}
\affiliation{Theoretical Physics III, Center for Electronic Correlations and Magnetism, University of Augsburg, 86135 Augsburg, Germany}

\begin{abstract}
A recently introduced one-dimensional two-particle Bose-Hubbard model with a single impurity [J.~M. Zhang \textit{et al}., Phys.~Rev.~Lett. \textbf{109}, 116405 (2012)]
is studied on {\it finite} lattices. The model possesses a discrete reflection symmetry and we demonstrate that all eigenstates odd under this symmetry can be obtained with a generalized Bethe ansatz if periodic boundary conditions are imposed. Furthermore, we provide numerical evidence that this holds true for open boundary conditions as well. The model exhibits backscattering at the impurity site---which usually destroys integrability---yet there exists an integrable subspace. We investigate the non-integrable even sector numerically and find a class of states which have almost the Bethe ansatz form. These weakly diffractive states correspond to a weak violation of the non-local Yang-Baxter relation which is satisfied in the odd sector. We bring up a method based on the Prony algorithm to check whether a numerically obtained wave function is in the Bethe form or not, and if so, to extract parameters from it. This technique is applicable to a wide variety of other lattice models. 
\end{abstract}

\maketitle

\section{Introduction}
Exactly solvable models are always appreciated, not only because of the 
beauty they embody but also because of the pivotal role they can play. Among the exactly solvable 
models known so far, many are solved by the Bethe 
ansatz for the wave function \cite{bethe}. A prerequisite for the consistency 
of this ansatz is the absence of diffraction \cite{sutherland}, which was 
first  explicitly  pointed out by McGuire \cite{mcguire}, and is formally 
characterized by the Yang-Baxter equation \cite{yang, baxter}.

Recently, in an investigation  motivated 
by atomic physics, the present authors discovered 
that a two-particle lattice model, 
whose continuum counterpart was previously 
claimed to be diffractive by McGuire \cite{mcguire},
has eigenstates in the Bethe ansatz form although they do not span the whole Hilbert space \cite{prl2012,pra2013}. The model is  very simple---it consists of two interacting identical bosons moving on a one-dimensional \textit{infinite} lattice with a single site defect. The model possesses a reflection symmetry (parity) and the states odd under parity conform to the Bethe ansatz. Its original purpose was to investigate how the competition or cooperation between the interaction and the defect potential will affect the formation of bound states in the system, as interesting consequences are known in a similar object, i.e., the negative hydrogen ion H$^-$ \cite{rau01}. It turns out that the model can exhibit two kinds of exotic bound state, i.e., the bound state in the continuum \cite{wigner} and the bound state at threshold \cite{zero,mattis,nieto}.

It was  convenient to consider the model on the infinite line because the bound state can be unambiguously discerned from the extended states belonging 
to the continuous spectrum.
The Yang-Baxter relation is then sufficient to prove the validity of the Bethe ansatz in the odd subspace \cite{pra2013}.
However, the situation is different on a finite lattice \cite{zvyagin}. The boundary conditions impose now restrictions on the allowed momenta and as the impurity potential generates backscattered waves which in all known cases inevitably spoil integrability, it was not clear 
whether meaningful Bethe ansatz equations can be derived 
to determine the discrete spectrum and if so, whether their solutions span the full subspace with odd parity. 

We shall address these questions in the following. Both can be answered in the affirmative for odd lattice size $M$ and periodic boundary conditions. However, the framework of the Bethe ansatz has to be generalized to incorporate the non-trivial backscattering at the impurity. The standard approach (see e.g. \cite{andrei}) defines the $S$-matrix 
as a relation between local amplitudes, whereas in our case the amplitudes have to be defined in a non-local way. 

This paper is organized as follows. 
In Sec.~\ref{defmodel} we define the model and outline the formalism of the generalized Bethe ansatz. We show that it allows for a natural inclusion of the reflection (parity) symmetry and thus explains at the same time the integrability of the 
odd sector and the failure of the ansatz in the even sector. The representation of periodic boundary conditions in the formalism is discussed in some detail. We derive the Bethe ansatz equations for this case.
The proof that their solutions span indeed the whole space is given in the appendix.  

The partial solvability of this model contradicts common lore about the possibility to use the Bethe ansatz for systems with impurities which are neither located at the edges \cite{sklyanin} nor have fine-tuned features which effectively remove the backscattering \cite{anjo}. In the present case, our primary evidence
comes from numerical data obtained by exact diagonalization. 
This is the subject of Sec.~\ref{numerics}. Here we study both open and periodic boundary conditions with three different methods to corroborate the fact that
the odd sector is always integrable while the even sector is not.  
In particular, the Bethe ansatz checking method in Sec.~\ref{algorithm} allows us to confirm that all the eigenstates in the odd subspace have the Bethe form.
On the other hand, the even sector exhibits eigenstates which are composed of only  a few Fourier components and thus resemble Bethe states.
These ``weakly diffractive'' states are presented in Sec.~\ref{fourier}. We show
in Sec.~\ref{YBE} that some of them correspond to a weak violation of the generalized Yang-Baxter equation, while 
the ordinary Yang-Baxter equation
is strongly violated.
Sec.~\ref{conclusions} summarizes the results and lays out some directions for future work.  

\section{The model on a finite lattice}\label{defmodel}
The model describes two spinless interacting bosons 
hopping on a lattice with a single site defect. On a one-dimensional lattice 
with an odd number $M=2M'+1$ sites, 
the Hamiltonian reads 
\beq\label{h}
\hat{H} \!=\!\sum_{i=-M'}^{M'} \left[-(\hat{a}_i^\dagger \hat{a}_{i+1}+ {\rm h.c.})+ 
\frac{U}{2} \hat{a}_i^\dagger \hat{a}_i^\dagger \hat{a}_{i} \hat{a}_{i}\right] \!+ V\hat{a}_0^\dagger \hat{a}_0. 
\eeq
We measure energy in units of the hopping integral; the parameters $U$ and $V$ are the strengths of on-site interaction and the defect potential at site 0, respectively.
A crucial property of \eqref{h} is its invariance under the 
reflection $\P\hat{a}_i\P=\hat{a}_{-i}$. 

The Hamiltonian \eqref{h} conserves particle number and we work in the two-particle subspace. 
To set up the Bethe ansatz, we use the first quantized form of \eqref{h}; the wave function reads $\p(x_1,x_2)$ with $ x_1,x_2$ being two integers, $-M'\le x_j\le M'$ and Bose symmetry requires $\p(x_1,x_2)= \p(x_2,x_1)$.
The Hamiltonian acts on $\p(x_1,x_2)$ as
\begin{align}\label{h2}
\hat{H} \p(x_1,x_2)&=\sum_{\Delta =\pm 1} -\left[\p(x_1+\Delta ,x_2)+ \p(x_1,x_2+\Delta ) \right ]\nonumber \\
 &+\left[V(\delta_{x_1,0}+ \delta_{x_2,0})+ U
\delta_{x_1,x_2} \right]\p(x_1,x_2). 
\end{align}
The group generated by $\P[\p](x_1,x_2)=\p(-x_1,-x_2)$ is $\Zz_2$ and a state $\p(x_1,x_2)$ will be called even [odd] under parity if it satisfies
$\p(x_1,x_2)=\p(-x_1,-x_2)$ [$\p(x_1,x_2)=-\p(-x_1,-x_2)$]. 

\subsection{Generalized Bethe ansatz}\label{framework}
\subsubsection{Non-local Yang-Baxter equation}
\begin{figure}[t]
\includegraphics[width= 0.3\textwidth ]{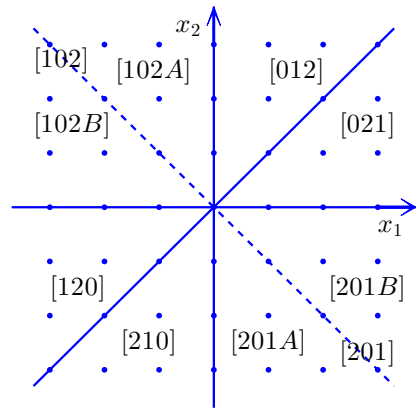}
\caption{The eight regions of the $x_1$-$x_2$ plane used in the generalized Bethe ansatz.  Regions $[102]$ and $[201]$ are split into two subregions although there is no interaction on the boundary between them (dashed line). This is the major difference between the formalism in this paper and the old formalism in Ref.~\cite{pra2013}. }
\label{lattice}
\end{figure}
The one-dimensional elastic scattering between the particles ($U$) leaves the momenta pairwise invariant but the impurity potential $V$ generates reflected waves with opposite momentum. Therefore only the absolute values of the momenta 
could serve as good quantum numbers. We try to generalize the Bethe ansatz, accounting for this simple type of diffraction through the introduction of an internal quantum number $\s_j=\pm 1$ for each particle which denotes the sign of the
corresponding momentum. In this way the scattering phases in a system of spinless bosons are replaced by non-commutative $S$-matrices. The generalized ansatz 
reads,
\beq
\p(x_1,x_2)=
{\cal S}\sum_{R, \s_1,\s_2}A^R_{\s_1\s_2}e^{i(k_1\s_1x_1+k_2\s_2x_2)}\chi_R(x_1,x_2).
\label{ansatz}
\eeq
The wavefunction is composed from plane waves defined in each region $R=[ijk]$
with $i,j,k\in\{0,1,2\}$ denoting the set $\{x_1,x_2\}$ in the
$x_1$-$x_2$ plane with
\beq
[ijk] \quad \leftrightarrow \quad x_i \le x_j \le x_k
\label{regions1}
\eeq
and $x_0=0$. $\chi_R(x_1,x_2)$ is the characteristic function of region $R$
and $\cal S$ symmetrizes the wave function with respect to $x_1$ and $x_2$.
The pseudo-spin index $\s_j\in\{+1,-1\}$ accounts for the possible backscattering at the impurity site. 
Due to the interactions at the boundaries of the regions $R$, the amplitudes $A^R_{\s_1,\s_2}$ are in general different
in the six regions given by \eqref{regions1}. 
For example, the amplitudes $A^{[021]}_{\s_1\s_2}$ and   $A^{[012]}_{\s_1\s_2}$ are related by the $\s_j$-dependent scattering phase  ($s_i= \sin k_i$, $i=1,2$)
\beq
A^{[012]}_{\s_1\s_2}=\frac{\s_1 s_1-\s_2 s_2 +i\frac{U}{2}}
{\s_1  s_1-\s_2  s_2 -i\frac{U}{2}}A^{[021]}_{\s_1\s_2}.
\label{s-phase}
\eeq
The scattering phases \eqref{s-phase} are unimodular for real momenta $k_j$.
On the other hand, the amplitudes $A^{[012]}_{++}, A^{[012]}_{-+}$ are related to
$A^{[102]}_{++}, A^{[102]}_{-+}$ as
\beq
\left(\!
\begin{array}{c}
A^{[102]}_{++}\\
A^{[102]}_{-+}
\end{array}
\!\right)
=
\left(\!
\begin{array}{cc}
1+\frac{iV}{2s_1} & \frac{iV}{2s_1}\\
 -\frac{iV}{2s_1} & 1-\frac{iV}{2s_1}
\end{array}
\!\right)
\left(\!
\begin{array}{c}
A^{[012]}_{++}\\
A^{[012]}_{-+}
\end{array}
\!\right).
\label{simpold}
\eeq
The matrix appearing in \eqref{simpold} maps the amplitudes with particle 1
on the right of the impurity to the amplitudes corresponding to particle 1 being on the left of the impurity. This entails that it is {\it not} unitary but preserves instead the particle current across the impurity:
\beq
|A^{[102]}_{++}|^2-|A^{[102]}_{-+}|^2=
|A^{[012]}_{++}|^2-|A^{[012]}_{-+}|^2.
\label{current}
\eeq
This non-unitarity is the main obstacle to derive the Bethe ansatz equations even if the ansatz \eqref{ansatz} should be consistent - which is not the case in general, the reason being the partial transmission/reflection generated by the impurity as exemplified in \eqref{simpold}. It turns out that the parity symmetry of the model is able to restore integrability to a certain degree in the two-particle sector. To see that, it is mandatory to define the amplitude vectors in such a way that all $S$-matrices become unitary. 

\begin{figure}[t!]
\centering
\vspace*{0.5cm}
\includegraphics[width=0.8\columnwidth]{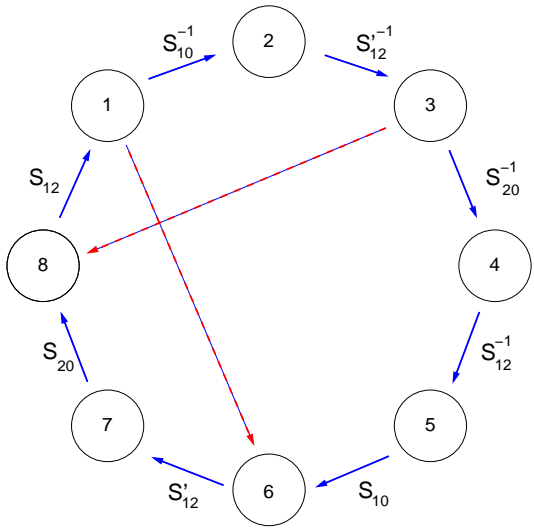}
\caption{(Color online) The eight amplitudes connected by $S$-matrices derived from the interaction terms in the Hamiltonian \eqref{h2}. The red lines correspond to transfer of particle 1 (from amplitude 3 to 8) 
resp. particle 2 (from amplitude 1 to 6) around the ring with periodic boundary conditions assumed. 
}
\label{octogon}
\end{figure}

All scattering processes are necessarily unitary if considered not as a transfer from ``right to left'' as in \eqref{simpold} but from ``past to future'':
One relates 
the set $A_{++}^{[102]},A_{-+}^{[012]}$ to the set  $A_{++}^{[012]},A_{-+}^{[102]}$, i.e. the incoming to the outgoing waves,
 \beq
\left(\!
\begin{array}{c}
A^{[012]}_{++}\\
A^{[102]}_{-+}
\end{array}
\!\right)
=
\frac{1}{s_1 +i\frac{V}{2}}\left(\!
\begin{array}{cc}
s_1 & -i\frac{V}{2}\\
 -i\frac{V}{2} & s_1
\end{array}
\!\right)
\left(\!
\begin{array}{c}
A^{[102]}_{++}\\
A^{[012]}_{-+}
\end{array}
\!\right).
\label{simpnew}
\eeq
The matrix in \eqref{simpnew} is unitary for real $k_1$ as it should be.
However, it collects amplitutes belonging to different regions into a single vector, in contrast to \eqref{simpold}. A consistent formalism has to express the scattering between particle 1 and 2 with non-local amplitudes as well which makes it convenient to split the regions $[102]$ and $[201]$ into two subregions
(see Fig.~\ref{lattice}),
\begin{align}
[102A] \; \leftrightarrow \; x_1 \le 0 \le x_2, &\quad |x_1| \le |x_2|, \nn\\
[102B] \; \leftrightarrow \; x_1 \le 0 \le x_2, &\quad |x_1| \ge |x_2|, \nn\\
[201A] \; \leftrightarrow \; x_2 \le 0 \le x_1, &\quad |x_1| \le |x_2|, \\
[201B] \; \leftrightarrow \; x_2 \le 0 \le x_1, &\quad |x_1| \ge |x_2|. \nn
\end{align}
The interaction part of the Hamiltonian vanishes on the boundary between
$[j0kA]$ and $[j0kB]$, so the corresponding amplitudes have to be the same.
This will be guaranteed if the $S$-matrices are chosen correctly.
We collect the local amplitudes $A^R_{\s_1\s_2}$ into the following eight 
four-vectors,
\begin{align}
\A_1=(A^{[012]}_{--},A^{[201A]}_{-+},A^{[102A]}_{+-},A^{[210]}_{++})^T,\nn\\
\A_2=(A^{[012]}_{+-},A^{[201A]}_{++},A^{[102A]}_{--},A^{[210]}_{-+})^T,\nn\\
\A_3=(A^{[021]}_{+-},A^{[201B]}_{++},A^{[102B]}_{--},A^{[120]}_{-+})^T,\nn\\
\A_4=(A^{[021]}_{++},A^{[201B]}_{+-},A^{[102B]}_{-+},A^{[120]}_{--})^T,\nn\\
\A_5=(A^{[012]}_{++},A^{[201A]}_{+-},A^{[102A]}_{-+},A^{[210]}_{--})^T,\\
\A_6=(A^{[012]}_{-+},A^{[201A]}_{--},A^{[102A]}_{++},A^{[210]}_{+-})^T,\nn\\
\A_7=(A^{[021]}_{-+},A^{[201B]}_{--},A^{[102B]}_{++},A^{[120]}_{+-})^T,\nn\\
\A_8=(A^{[021]}_{--},A^{[201B]}_{-+},A^{[102B]}_{+-},A^{[120]}_{++})^T.\nn
\end{align}
The wavefunction reads in terms of the components $A_j^k$, ($j=1\ldots 8$, 
$k=1\ldots 4$),
\begin{align}
\p&(x_1,x_2)= \nn\\
{\cal S}\Big\{ &\Big(\sum_k[A_8^k\chi_{A_8^k}
+A_1^k\chi_{A_1^k}]\Big)e^{-i(k_1|x_1|+k_2|x_2|)}\nn\\
+& \Big(\sum_{k}[A_2^k\chi_{A_2^k}
+A_3^k\chi_{A_3^k}]\Big)e^{ik_1|x_1|-k_2|x_2|}\nn\\
+&\Big(\sum_k[A_4^k\chi_{A_4^k}
+A_5^k\chi_{A_5^k}]\Big)e^{i(k_1|x_1|+k_2|x_2|)} \label{ansatz2}\\
+& \Big(\sum_{k}[A_6^k\chi_{A_6^k}
+A_7^k\chi_{A_7^k}]\Big)e^{-ik_1|x_1|+k_2|x_2|}\Big\},\nn
\end{align}
where $\chi_{A_j^k}=\chi_{R(A_j^k)}(x_1,x_2)$ denotes the characteristic function of the region belonging to $A_j^k$. Note that the exponential factor is invariant under $\P$ and the same for all components $A_j^k$ of $\A_j$. Moreover, we have for all $j=1\ldots 8$ the relation
\beq
\P[\chi_{A_j^k}](x_1,x_2)=\chi_{A_j^{5-k}}(x_1,x_2).
\eeq
It follows that
\beq
\P[\p](x_1,x_2;\{A_j^k\})=
\p(x_1,x_2;\{A_j^{5-k}\}),
\eeq
which means that $\P$ acts on $\p(x_1,x_2)$ by transforming each amplitude vector $\A_j$ with the matrix $R_{kl}=\d_{k,5-l}$, $k,l=1,\ldots,4$. 
The projection of $\p(x_1,x_2)$ onto the even (odd) subspace  is done by projecting the amplitudes $\A_j$ onto the two-dimensional eigenspace of $\hat{R}$ with eigenvalue $+1$ ($-1$). The  $S$-matrices $\S(j,l)=\S^{-1}(l,j)$ for $1\le j,k\le 8$ connect ``adjacent'' amplitudes $\A_j=\S(j,j\pm 1)\A_{j\pm 1}$ ($8+1=1$ because $\A_8$ is adjacent to $\A_1$).
$\S(2,3),\S(4,5),\S(6,7),\S(8,1)$ correspond to an exchange of particle 1 with 2
and $\S(1,2),\S(5,6)$ [$\S(3,4),\S(7,8)$] to scattering of particle 1 [2] at the impurity, see Fig.~\ref{octogon}.

The determination of the $S$-matrices from \eqref{h2} is straightforward \cite{pra2013}. Defining
\beq
\a=\frac{s_1-s_2+i\frac{U}{2}}{s_1 -s_2 -i\frac{U}{2}},
\quad
\b=\frac{s_1+s_2+i\frac{U}{2}}{s_1 +s_2 -i\frac{U}{2}},
\eeq
we find e.g. for $\hat{S}(1,8)$,
\beq
\S(1,8)=
\left(
\begin{array}{cccc}
\a^{-1}&0&0&0\\
0&1&0&0\\
0&0&1&0\\
0&0&0&\a^{-1}
\end{array}
\right),
\eeq
and for $\hat{S}(2,1)$,
\beq
\S(2,1)=
\frac{1}{s_1 +i\frac{V}{2}}\left(
\begin{array}{cccc}
-i\frac{V}{2}&0& s_1 &0\\
0&-i\frac{V}{2}&0& s_1\\
s_1&0&-i\frac{V}{2}&0\\
0& s_1& 0 &-i\frac{V}{2}
\end{array}
\right).
\eeq
Each amplitude can be obtained from any other by following a path connecting
them on the circle depicted in Fig.~\ref{octogon}. 
The ansatz \eqref{ansatz}, \eqref{ansatz2} will be consistent if the amplitudes
$\A_j$ are not overdetermined by the $S$-matrices.
The amplitude $\A_5$ is given by $\A_1$ in two ways,
\begin{subequations}
\begin{eqnarray}
\A_5 &=& \S(5,4)\S(4,3)\S(3,2)\S(2,1)\A_1, \\
\A_5 &=& \S(5,6)\S(6,7)\S(7,8)\S(8,1)\A_1.
\end{eqnarray}
\end{subequations}
Because of the relations 
\begin{align}
\S(8,1)&=\S(5,4), \quad \S(7,8)=\S(4,3), \nn\\
\S(6,7)&=\S(3,2), \quad \S(5,6)=\S(2,1),
\end{align}
we obtain the analogue of the Yang-Baxter equation (YBE),
\begin{eqnarray}
& \S(8,1)\S(7,8)\S(6,7)\S(5,6)  \nonumber \\
& =\S(5,6)\S(6,7)\S(7,8)\S(8,1). \label{YBE1}
\end{eqnarray}
This generalized consistency relation is quadrilinear instead of trilinear 
in the $S$-matrices. The reason is the existence of two different $S$-matrices describing the scattering between particle 1 and 2: Although both $\S(1,8)$
and $\S(7,6)$ correspond to the exchange of the two particles, they are not identical; if we define $\S(1,8)=\S_{12}$ and $\S(7,6)=\S_{12}'$ together with
$\S(8,7)=\S_{20}$, $\S(1,2)=\S_{10}$, we obtain the
generalized YBE in the following, more familiar form,
\beq
\S_{10}\S_{12}'\S_{20}\S_{12}=\S_{12}\S_{20}\S_{12}'\S_{10}.
\label{YBE2}
\eeq
This relation is formally equivalent to Sklyanin's condition for integrable boundary terms \cite{sklyanin}; the difference lies in the physical interpretation: 
in the systems considered by Sklyanin, the particles are only reflected but not transmitted by the boundary potential (the amplitudes are local). 
Due to the parity invariance of \eqref{h}, the $\S$-matrices are block-diagonal in the eigenbasis of $\P$. 
One finds  for the even sector,
\begin{align}
\S^+_{10}=&\frac{1}{s_1-i\Vh}\left(\!\begin{array}{cc}
i\Vh & s_1\\
s_1 & i\Vh
\end{array}\!\right),
{\S}^{\prime +}_{12}=\left(\!\begin{array}{cc}
\b & 0\\
0 & 1
\end{array}\!\right), \nn\\
\S^+_{20}=&\frac{1}{s_2-i\Vh}\left(\!\begin{array}{cc}
i\Vh & s_2\\
s_2 & i\Vh
\end{array}\!\right),
\S^+_{12}=\left(\!\begin{array}{cc}
\alpha^{-1} & 0\\
0 & 1
\end{array}\!\right), 
\end{align}
and for the odd sector,
\begin{align}
\S^-_{10}=&\frac{-1}{s_1-i\Vh}\left(\!\begin{array}{cc}
-i\Vh & s_1\\
s_1 & -i\Vh
\end{array}\!\right),
{\S}^{\prime -}_{12}=\left(\!\begin{array}{cc}
1 & 0\\
0 & \b
\end{array}\!\right), \nn\\
\S^-_{20}=&\frac{1}{s_2-i\Vh}\left(\!\begin{array}{cc}
i\Vh & s_2\\
s_2 & i\Vh
\end{array}\!\right),
\S^-_{12}=\left(\!\begin{array}{cc}
1 & 0\\
0 & \a^{-1}
\end{array}\!\right). \label{smat}
\end{align}
One checks now that \eqref{YBE2} is violated for
$\sin k_1\sin k_2 UV\neq 0$ in the even sector but satisfied identically in
$k_j,U,V$ in the odd sector. The ansatz \eqref{ansatz} is therefore consistent for parity-odd states but fails in the even sector (see Sec.~\ref{YBE}).

\subsubsection{Bethe ansatz equations}

We confine the analysis from now on to the odd sector. The momenta $k_1,k_2$ are quantized by the boundary conditions (BC). For periodic BC we have
$\p(x_1,x_2)=\p(x_1+M,x_2)=\p(x_1,x_2+M)$. Open BC correspond to
$\p(\pm(M'+1),x_2)=\p(x_1,\pm(M'+1))=0$. We shall now demonstrate how periodic BC can be implemented within our non-local framework, leaving open BC for future study (they are treated numerically in Sec.~\ref{numerics}).   

Because of the non-local nature of the amplitudes $\A_j$, a given particle (say particle 1) moves in both directions simultaneously if the regions are connected via the $S$-matrices $\S(j_1,j_2)$. It is therefore convenient to treat each component $A_j^k$ separately, before projection onto the odd subspace. For example, the amplitude $A_1^1$ describes particle 1 with negative momentum $-k_1$ to the right of the impurity, $0<x_1<x_2<M'$. $\S(2,1)$ maps it to $A_2^3$ to the left of site 0. $\S(3,2)$ corresponds to crossing the border between $[102A]$ and $[102B]$, leading to $A_3^3$, of course without changing the value of amplitude $A_2^3$. Now particle 1 is close to the left border of the system, periodicity for values $x_1<-M'$ means proportionality of $A_3^3$ to ${A_3^3}'$ with $0<x_2<x_1<M'$. 
We have
\beq
\p \propto A_3^3e^{i(-k_1)(-M'-x)}={A_3^3}'e^{i(-k_1)(M'-x+1)},    
\eeq
or ${A_3^3}'=e^{ik_1M}A_3^3$. Now the configuration of ${A_3^3}'$ can be recognized as  that of $A_8^1$, which is mapped by $\S(1,8)$ back onto $A_1^1$.
Similarly, $A_1^3$ describes a right-moving particle with positive momentum $k_1$, which is transferred by $\S(2,1)$ to $A_2^1$, and in turn by $\S(3,2)$ to $A_3^1$. This time the exchange between the two particles leads to a non-trivial phase factor, because the border between $[012]$ and $[021]$ is crossed. The regions
$A_3^1$ and ${A_3^1}'$ with $-M'<x_1<0<x_2$ are related by periodicity as
\beq
\p\propto A_3^1e^{ik_1(M'+x)}={A_3^1}'e^{ik_1(-M'-1+x)},
\eeq 
meaning ${A_3^1}'=e^{ik_1M}A_3^1$. ${A_3^1}'$ corresponds to $A_8^3$, mapped by 
$\S(1,8)$ onto the original amplitude $A_1^3$. 
The phase factor $e^{ik_1M}$ picked up during transfer from $\A_3$ to $\A_8$ (see the red arrows in Fig.~\ref{octogon}) is independent from the sign of the momenta.
Collecting all amplitudes and projecting onto the odd subspace, we conclude that  
the odd part of $\A_1$ must be an eigenvector of
\beq
 \M_1=\frac{1}{s_1+i\Vh}
\left(\!\begin{array}{cc}  
s_1\b^{-1} & i\Vh\b^{-1}\\
i\Vh\a^{-1} & s_1\a^{-1}
\end{array}
\!\right)
\label{matrix1}
\eeq 
with eigenvalue $e^{-ik_1M}$. In an analogous way, by transporting particle 2 around the ring, one obtains that the odd part of $\A_1$ is an eigenvector
of
\beq
\M_2=\frac{1}{s_2+i\Vh}
\left(\!\begin{array}{cc}  
s_2\b^{-1} & -i\Vh\a\b^{-1}\\
-i\Vh & s_2\a
\end{array}
\!\right)
\label{matrix2}
\eeq
with eigenvalue $e^{-ik_2M}$.
Indeed, $[\M_1,\M_2]=0$, so both matrices are simultaneously diagonalizable. Moreover, they are unitary for real $k_1,k_2$. Let us denote their eigenbasis as $\{\v_+,\v_-\}$. The corresponding eigenvalues of 
$\M_1$ and $\M_2$ read
\begin{widetext}
\begin{subequations}\label{lam12}
\begin{align}
\lo_\pm=&\frac{s_1(s_1^2-s_2^2+\frac{U^2}{4})\pm i\left[(U^2-V^2)s_1^2s_2^2+\frac{V^2}{4}
(s_1^2+s_2^2+\frac{U^2}{4})^2\right]^{1/2}}{s_1(s_1^2-s_2^2-\frac{U^2}{4}
-\frac{UV}{2})+i\left[\frac{V}{2}
(s_1^2-s_2^2-\frac{U^2}{4})+Us_1^2\right]} ,\\
\lz_\pm=&\frac{s_2(s_2^2-s_1^2+\frac{U^2}{4})\pm 
i\left[(U^2-V^2)s_1^2s_2^2+\frac{V^2}{4}
(s_1^2+s_2^2+\frac{U^2}{4})^2\right]^{1/2}}{s_2(s_2^2-s_1^2
-\frac{U^2}{4}-\frac{UV}{2})
+i\left[\frac{V}{2}
(s_2^2-s_1^2-\frac{U^2}{4})+Us_2^2\right]}. 
\end{align}
\end{subequations}
\end{widetext}
One notes that the eigenvalue of $\M_2$ obtains from the corresponding eigenvalue of $\M_1$ by exchange of $k_1$ and $k_2$, as expected. 
If one writes
\beq
\l^{(j)}_\pm=\frac{a_j\pm ib}{c_j+id_j},
\eeq
we have
\beq 
a_j^2+b^2=c_j^2+d_j^2,
\label{identity}
\eeq
which guarantees unimodularity of $\l^{(j)}_\pm$ for real $k_1,k_2$. The Bethe ansatz equations read 
\beq
e^{-ik_1M}=\lo_\pm(k_1,k_2),\quad e^{-ik_2M}=\lz_\pm(k_1,k_2),
\label{BAE1}
\eeq
which are coupled equations for $k_1$ and $k_2$. In general both $\v_+$ 
and $\v_-$ yield solutions obtained from $\{\lo_+,\lz_+\}$ and  $\{\lo_-,\lz_-\}$, respectively. 

An interesting detail can be read off from \eqref{lam12} immediately: 
if $U=-V$ and $U>0$ ($U<0$) the eigenvalues
$\l^{(j)}_+$ ($\l^{(j)}_-$) become 1 for all $k_1,k_2$. 
This means that all states in the subspaces
$\v_+$ ($\v_-$) have momenta quantized in multiples of $2\pi/M$ as in the non-interacting system. The effects of the impurity and the interaction (being equal but with opposite sign) compensate each other in one of the invariant subspaces
$\v_\pm$ and the particles behave in this subspace as if they were free bosons on a ring with $M$ sites.  

Although \eqref{BAE1} has formally the same structure as the Bethe ansatz equations in the two-particle sector of known integrable systems, 
the right-hand sides are much more complicated, leading to more types of possible solutions. It turns out that there are not just real 
solutions and strings \cite{andrei} but two additional types, which we shall discuss in detail in the next section, presenting a numerical analysis of the eigenstates. 

\section{Numerical analysis in three aspects}\label{numerics}
 In this section we study the model via exact diagonalization on finite lattices, for both open and periodic BC. We shall use three methods to analyze the numerical data. These will corroborate the results of 
Sec.~\ref{framework} and provide numerical evidence that the odd sector is integrable regardless of the boundary condition or the parity of the lattice size.

\subsection{Spectral graph}\label{spectrumgraph}
\begin{figure*}[tb]
\begin{minipage}[b]{0.945 \textwidth}
\includegraphics[width=\textwidth]{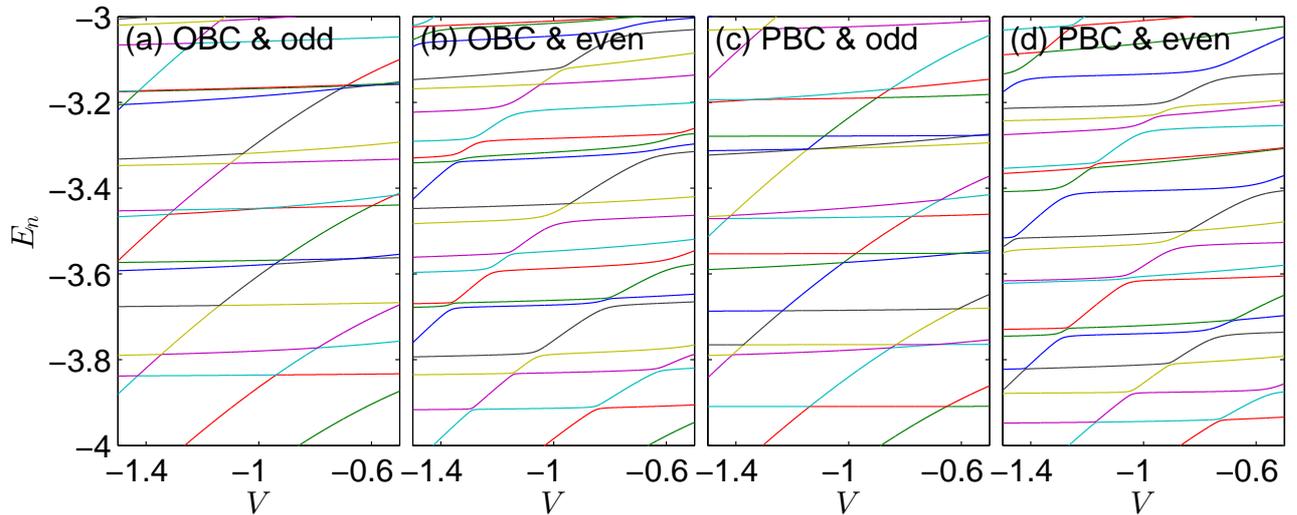}
\end{minipage}
\caption{\label{spectrum}(Color online) Spectral graph of the model in odd or even parity subspaces and with open boundary condition (OBC) or periodic boundary condition (PBC). In all the four panels, $M=31$ and the on-site interaction $U=2$. Level crossing in (a) and (c), and level anti-crossing in (b) and (d), are apparent.}
\end{figure*}

The first evidence that the odd-parity subspace is integrable while the even one is not comes from the behavior of the energy spectrum of the model as a function of $V$ (or $U$ as well). 

In Fig.~\ref{spectrum}, we plot the spectrum of the model as a function of $V$ in the odd and even parity subspaces and with open or periodic boundary conditions. We see that regardless of the boundary condition, levels in the odd subspace cross each other while levels in the even subspace repel each other. These two different behaviors strongly hint on the integrability of the odd subspace and the nonintegrability of the even subspace \cite{stepanov}. 

\subsection{Bethe-form checking}\label{algorithm}
\begin{figure*}[tb]
\begin{minipage}[b]{0.95\textwidth}
\includegraphics[width=\textwidth]{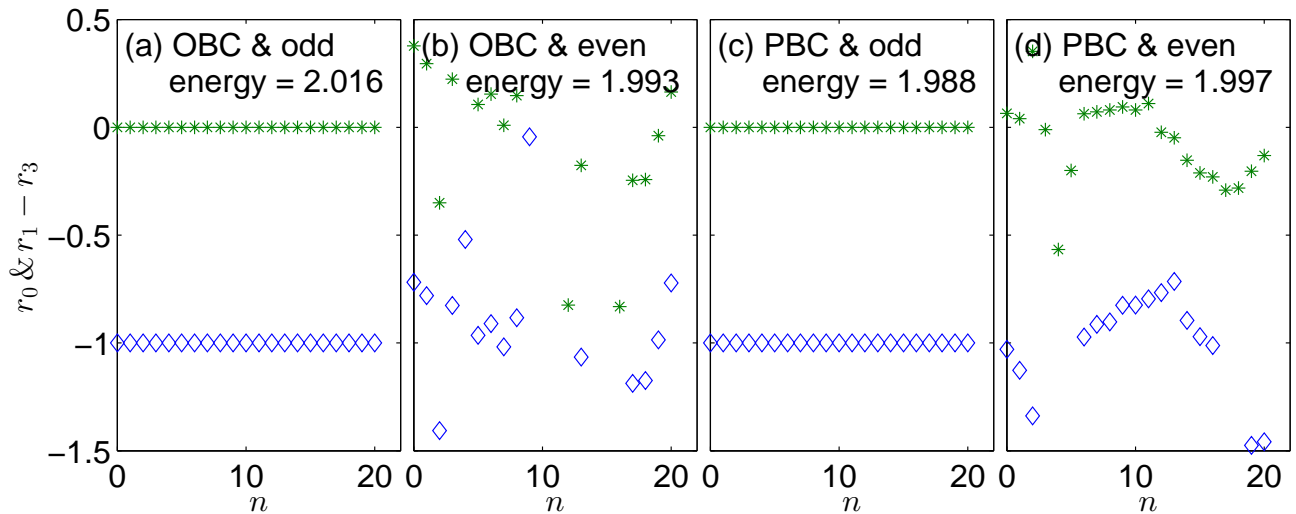}
\end{minipage}
\caption{\label{check}(Color online) Checking whether an eigenstate obtained by numerical exact diagonalization has the Bethe ansatz form or not by studying whether $r_0 $ (diamond $\diamond$) and $r_1- r_3 $ (asterisk $\ast$) are independent of $n$. In all the four panels, the lattice size is $M=55$ and $(V,U)= (-2,-2) $. In each panel, the boundary condition, the parity of the investigated eigenstate, and the energy of the state, are shown at the top. Note that in each panel, we choose the eigenstate whose energy is the closest to 2. We see that in (a) and (c), $r_0=-1$ and $r_1-r_3=0$ independent of $n$ as expected for a Bethe state, while in (b) and (d), $r_0$ and $r_1-r_3$ fluctuate strongly. Note also that in the later case, some data points are beyond the range of the vertical axis and thus missing on the panel.}
\end{figure*}

By full exact diagonalization of the Hamiltonian, we can obtain numerical values of all the eigenvalues and eigenstates of the model. A natural question is then whether it is possible to demonstrate that a wave function, whose value at each site $(x_1,x_2)$ is known,  can or cannot be decomposed into the form of (\ref{ansatz}). 
To answer this question, first we note that, if the wave function has this form, its 
values on a line which lies entirely in a particular region (say, the line $x_1=0$, $ x_2\geq 0$ in region $[102]$), is a superposition of at most four exponentials (if the four wave vectors $\pm k_1$ and $\pm k_2$ are all different, as is demonstrated below). Therefore, first we need to check whether the wave function evaluated on this line is a superposition of four exponentials.

\subsubsection{Prony's algorithm}
Fortunately, there exists a beautiful algorithm, namely Prony's algorithm \cite{prony}, which can be used to check whether a function is a superposition of several exponentials and if it is, to extract the exponents.
The logic of this algorithm is as follows. Let a function $n\rightarrow g_n $ defined on the integers be a superposition of four exponentials,
\begin{equation}\label{g}
g_n= w_1 e^{c_1n }+ w_2 e^{c_2 n} + w_3 e^{c_3 n}+ w_4 e^{c_4 n},
\end{equation}
where the exponents $c$'s are assumed to be all different, and the $w$'s are all constant coefficients. Here we take four exponentials merely because of the context---the algorithm works for an arbitrary number of exponentials.
Because the $c$'s are all different, the Vandermonde matrix 
\begin{equation}
K=\left(
\begin{array}{cccc}
1 & e^{c_1} & e^{2 c_1} & e^{3 c_1} \\
 1 & e^{c_2} & e^{2 c_2} & e^{3 c_2} \\
 1 & e^{c_3} & e^{2 c_3} & e^{3 c_3} \\
 1 & e^{c_4} & e^{2 c_4} & e^{3 c_4}
\end{array}
\right)
\end{equation}
is non-singular and the linear equation 
\begin{eqnarray}\label{eqforq}
\left(
\begin{array}{cccc}
1 & e^{c_1} & e^{2 c_1} & e^{3 c_1} \\
 1 & e^{c_2} & e^{2 c_2} & e^{3 c_2} \\
 1 & e^{c_3} & e^{2 c_3} & e^{3 c_3} \\
 1 & e^{c_4} & e^{2 c_4} & e^{3 c_4}
\end{array}
\right) \left( \begin{array}{c}r_0 \\ r_1 \\r_2 \\r_3
\end{array} \right)= \left( \begin{array}{c} e^{4c_1}\\e^{4c_2}\\e^{4c_3}\\e^{4c_4}
\end{array}\right)
\end{eqnarray}
has a unique solution for the $r$'s. It is readily seen that this fact implies the linear iterative relation
\begin{eqnarray}\label{iterative}
g_{n+4}= r_3 g_{n+3}+ r_2 g_{n+2}+ r_1 g_{n+1}+ r_0 g_{n},
\end{eqnarray} 
which holds regardless of the values of the $w$'s. 

Now consider such a linear equation for the $z$'s (note that the $4\times 4$ matrix on the left hand side is a Hankel matrix),
\begin{eqnarray}\label{line}
\left( \begin{array}{cccc}
g_n & g_{n+1} & g_{n+2} & g_{n+3}\\
g_{n+1} & g_{n+2} & g_{n+3} & g_{n+4} \\
g_{n+2} & g_{n+3} & g_{n+4} &g_{n+5}\\
g_{n+3} &g_{n+4} &g_{n+5} &g_{n+6} 
\end{array}   \right)\left( \begin{array}{c} z_0 \\z_1 \\z_2 \\z_3
\end{array}\right)= \left( \begin{array}{c}
g_{n+4}\\g_{n+5} \\g_{n+6} \\g_{n+7}
\end{array} \right).\quad 
\end{eqnarray}
If the value of the function $g$ is known at each integer $n$, the $4\times 4$ matrix on the left hand side and the $4 \times 1$ vector on the right hand side can be constructed, and the linear equation is well-posed. Now the point is that, in view of (\ref{iterative}), the solution $(z_0,z_1,z_2,z_3)^T$ would be equal to $(r_0,r_1,r_2,r_3)^T$ if $g$ has the form (\ref{g}). That is, it is a constant vector \textit{independent} of $n$.
This is a necessary condition for the function $g$ to have the form (\ref{g}).

Therefore, by constructing and solving the linear equation in (\ref{line}) at different $n$, and observing whether the solution varies with $n$, one can determine whether a function $g$ has the form (\ref{g}) or not. If it is, which means the values of the $r$'s have been determined, one has at least two options to determine the values of the $c$'s. One can either solve them inversely by using Eq.~(\ref{eqforq}) as we will do below, or by diagonalizing the $4\times 4$ transfer matrix in the following equation
\begin{eqnarray}\label{eigens}
\left( \begin{array}{c}
g_{n+4}\\g_{n+3}\\g_{n+2}\\g_{n+1}
\end{array} \right) = \left( \begin{array}{cccc}
r_3 & r_2 & r_1 & r_0 \\
1 & 0 & 0 & 0 \\
0 & 1 & 0 & 0 \\
0 & 0 & 1 & 0 
\end{array} \right) \left( \begin{array}{c}
g_{n+3}\\g_{n+2}\\g_{n+1}\\g_{n}
\end{array} \right) ,
\end{eqnarray}
which is a reformulation of the iterative relation  (\ref{iterative}).

Now let us come back to our problem. Suppose we take the line with $x_1=0$ and $x_2\geq 0$, and consider the wave function on this line as a function of $x_2$. 
The range of $x_2$ depends on the boundary condition. In the case of open BC, we have to confine ourself to region $[102]$ and $x_2$ can take values from 0 up to $M'$; while in the case of periodic BC, we can advance into region $[210]$ and $x_2$ can take values from 0 up to $M-1$. We can then use Prony's algorithm above to check whether the wave function has the Bethe ansatz form or not. 

In our specific problem, the $c$'s are a permutation of $(ik_1,-ik_1, ik_2 ,-ik_2)$. It is tedious but straightforward to solve the $r$'s in Eq.~(\ref{eqforq}) as 
\begin{subequations}\label{qs}
\begin{eqnarray}
r_0 & =& -1,  \\
r_1 & =& e^{ik_1}+ e^{-ik_1}+ e^{ik_2}+ e^{-ik_2}=- E(k_1,k_2), \quad  \\
r_2 & =& -(e^{ik_1}+ e^{-ik_1})(e^{ik_2}+ e^{-ik_2})-2, \\
r_3 &= & r_1.
\end{eqnarray}
\end{subequations} 
The fact that $r_0=-1$ regardless of the values of $k_{1,2}$ is a stringent condition for the wave function to have the Bethe form. Besides this condition, $r_1 = r_3=-E$ is another very stringent one. Once these two conditions are verified, it is likely that the wave function has indeed the Bethe form. In Fig.~\ref{check}, we show $r_0$ and $r_1-r_3$ as functions of $x_2$ in different subspaces and under different boundary conditions. The lattice has $M=55$ sites and the state in each panel is sampled unbiased as the one whose energy is the closest to 2. We see that, for the odd-parity states, $r_0$ and $r_1-r_3$ are indeed independent of the position $x_2$ and take the values expected. On the contrary, for the even-parity states, they both fluctuate significantly, which rules out the possibility that the corresponding wave function has the Bethe form. 

By Eqs.~(\ref{qs}b) and (\ref{qs}c), we can solve from the values of $r_{1,2}$ the values of $\mu_j=e^{ik_j}+ e^{-ik_j}$ ($j=1,2$), from which in turn we can solve the wave vectors $k_{1,2}$. Finally, we can plug  $k_{1,2}$ into \eqref{ansatz}, solve for the amplitudes and compare the result with the one obtained by exact diagonalization. The inner product between them should be unity if they are the same.  

We have carried out this algorithm on lattices with sizes between 15 and 31 and with both kinds of boundary condition. We used the Multiple Precision Toolbox for MATLAB \cite{matlab} to solve for the eigenstates and eigenvalues to high precisions (say, with 40-60 digits) and then analyzed the eigenvectors one by one by using Prony's algorithm. In this way, we confirmed that all the odd-parity eigenstates have the form (\ref{ansatz}) while none of the even-parity eigenstates fit to it. 

Here some remarks are necessary. First, we need high precisions for the eigenstates because - as we shall see in the following - some eigenstates decay exponentially in at least one direction. If the precision is not high enough, the true values of the $g$'s in (\ref{line}) can be overwhelmed by the noise by orders of magnitude. Second, Prony's algorithm is very sensitive to any perturbation to the wave function. Specifically, for the weakly diffractive states in Figs.~\ref{moms} and \ref{strangemoms} below, we get $r$'s as random as in Figs.~\ref{check}(b) and \ref{check}(d). Therefore, the algorithm can only tell whether a state is \textit{exactly} in the Bethe form or not. It cannot distinguish weakly diffractive states from strongly diffractive states. 

\subsubsection{Four main categories of states}\label{3bands}

\begin{figure*}[tb]
\begin{minipage}[b]{0.95 \textwidth}
\includegraphics[width=\textwidth]{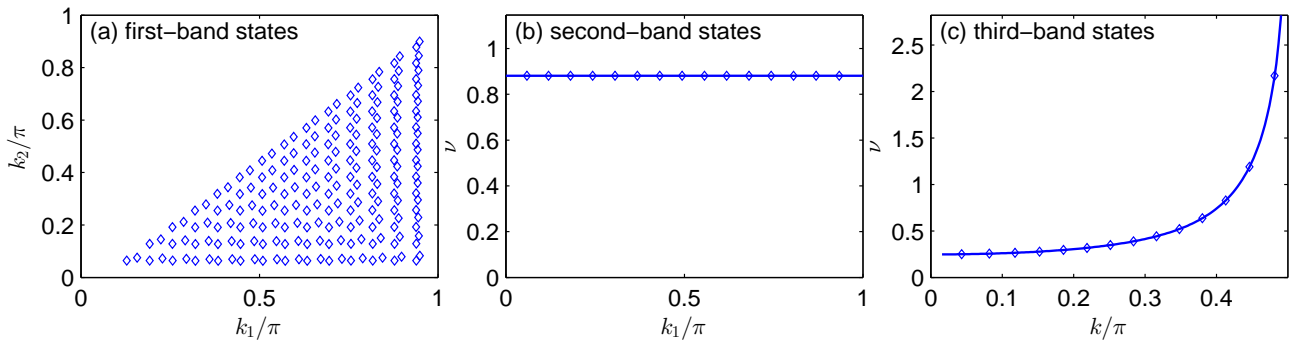}
\end{minipage}
\caption{\label{k1k2}(Color online) Classification of the wave vector pair $(k_1,k_2)$ for the odd-parity eigenstates in three primary categories. Each data point corresponds to a state. The parameters are $(V,U,M) = (-2, -1,31)$ and the boundary condition is open. In (a), $k_{1,2}$ are both real. The corresponding states belong to the first band. In (b), one wave vector is real while the other is purely imaginary, i.e., $(k_1,k_2)= (k_1, i\nu)$. The horizontal solid line indicates the value of $\nu_{imp}$ [see Eq.~(\ref{nuimp})]. The corresponding states belong to the second band. In (c), the two wave vectors are complex conjugates, i.e., $(k_1,k_2)= (k+i \nu, k-i \nu)$. The solid line indicates the function of (\ref{numol}). In (b) and (c), the deviations of the data points from the solid lines are too small to be visible on the current scale. By counting the number of states, we know that one state is missing on these three panels. It  belongs to category (iv) and is a bound state in the continuum \cite{prl2012, pra2013}.}
\end{figure*}

Previously it was shown that the model on the infinite line has three continuum bands with different nature \cite{prl2012, pra2013}. It turns out that in correspondence there are three main categories of $(k_1, k_2)$-pair in the odd sector, which are shown in the three panels of Fig.~\ref{k1k2} separately and explained below (because the wave function is real, the $r$'s are real and thus 
$\mu_{1,2} $ must be real simultaneously  or complex conjugate to each other): 

(i) $\mu_{1,2}$ are both real, and $\max \{ |\mu_1|, |\mu_2|\} \leq 2$. In this case, $k_{1,2}$ are both real and we can always choose them as $0\leq k_2\leq k_1 \leq \pi$ [see Fig.~\ref{k1k2}(a)]. The realness of $k_{1,2}$ means that the two particles are both delocalized on the lattice and they are not bound to each other. This is the feature of the first-band states \cite{prl2012, pra2013}. 

(ii) $\mu_{1,2}$ are both real, and $\min \{ |\mu_1|, |\mu_2|\} \leq 2 < \max \{ |\mu_1|, |\mu_2|\}$. In this case, one wave vector is real while the other is complex (depending on the sign of $V$, it can be purely imaginary or have a real part of $\pi$). We observe that [see Fig.~\ref{k1k2}(b)], the magnitude of the imaginary part can be well approximated by 
\begin{equation}\label{nuimp}
\nu_{imp}= \asinh(|V|/2) > 0.
\end{equation}
Here $\nu_{imp} $ is the inverse of the localization length of the defect mode induced by the impurity. This kind of $(k_1,k_2)$ indicates that one particle is localized by the impurity potential while the other is not, in accordance with the picture for the second band \cite{prl2012, pra2013}. 

(iii) $\mu_{1,2}$ are both complex, and $\mu_1^* = \mu_2$. As a result, $k_{1,2}$ are complex and can be chosen in the form $(k_1 ,k_2)= (k+i \nu, k-i \nu)$ with $\nu >0$ and $0<k<\pi $. It is observed that [see Fig.~\ref{k1k2}(c)], generally the relation
\begin{equation}\label{numol}
\nu =  \asinh(U/ (4 \cos k))
\end{equation}
holds approximately. This kind of complex $(k_1,k_2)$ with the relation (\ref{numol}) is what one expects for an interaction induced molecule state. Therefore, the corresponding state falls in the third band \cite{prl2012, pra2013}, where the two particles form a molecule and move on the lattice together. The approximate validity of \eqref{nuimp}, \eqref{numol} is derived in the Appendix using the Bethe ansatz equations \eqref{BAE1}.

We have thus confirmed the existence of three distinct continuum bands 
from another perspective. Note that the number of states in the first band is of the order  $M^2$, while those in the second and third bands are merely of the order  $M$. This is reasonable since in the latter two bands, always one degree of freedom is frozen. 

\begin{figure}[tb]
\includegraphics[width= 0.35\textwidth ]{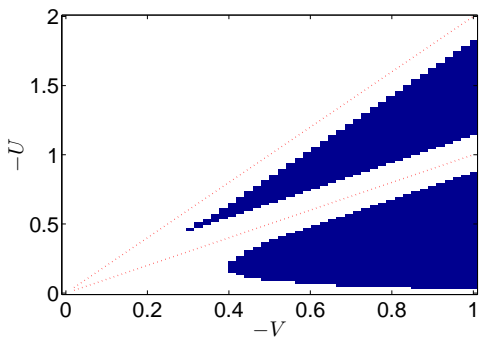}
\caption{(Color online) Region (the blue shaded part) where the category-(iv) state occurs, which would be an odd-parity bound state on an infinite lattice. The lattice has 31 sites and open boundary conditions. Due to the finiteness of the lattice size, the blue part is just a proper subset of the region $0< U/V <2$ where the infinite system shows an odd-parity bound state \cite{prl2012,pra2013}. 
\label{region}}
\end{figure}

Besides the three primary categories above, there is yet a fourth possibility, i.e., 

(iv) $\mu_{1,2}$ are both real, and $\min \{ |\mu_1|, |\mu_2|\}>2 $. In this case, $e^{ik_1}$ and $e^{ik_2}$ are both real. This possibility is related to the two analytically solvable odd-parity bound states on an infinite lattice \cite{prl2012,pra2013}, which appear alternatively in the regions $0<U/V<1$ and $1<U/V<2$. In fact, by our analytic analysis before \cite{prl2012,pra2013}, these two bound states have the property of $\min \{ |\mu_1|, |\mu_2|\}>2 $. Here it is observed that for given parameters $(V,U,M)$, this case occurs at most once. Moreover, due to the finiteness of the lattice size, the region where it occurs is a proper subset of the region $0< U/V< 2$. In Fig.~\ref{region}, the region where it occurs is shown for a $M=31$ lattice with open boundary conditions. 

We prove in the Appendix for periodic boundary conditions that the first band contains $(M-1)(M-3)/4$ states, the second band $M'$ states and the third band either $M'$ or $M' -1$ states, depending on the absence or presence
of a bound state. Altogether this are $(M^2-1)/4$ states, the full dimension of the odd subspace. 

\subsection{Momentum distribution}\label{fourier}
\begin{figure*}[thb]
\includegraphics[width=0.45\textwidth]{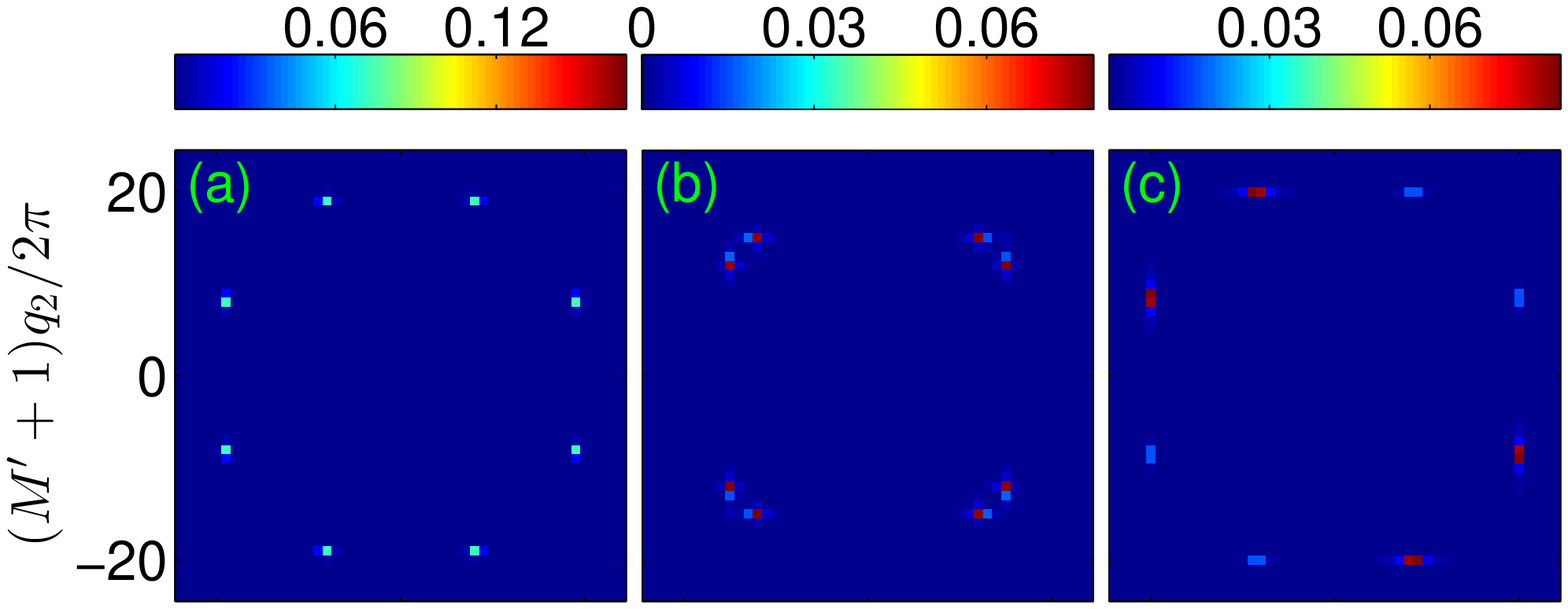}
\includegraphics[width=0.45\textwidth]{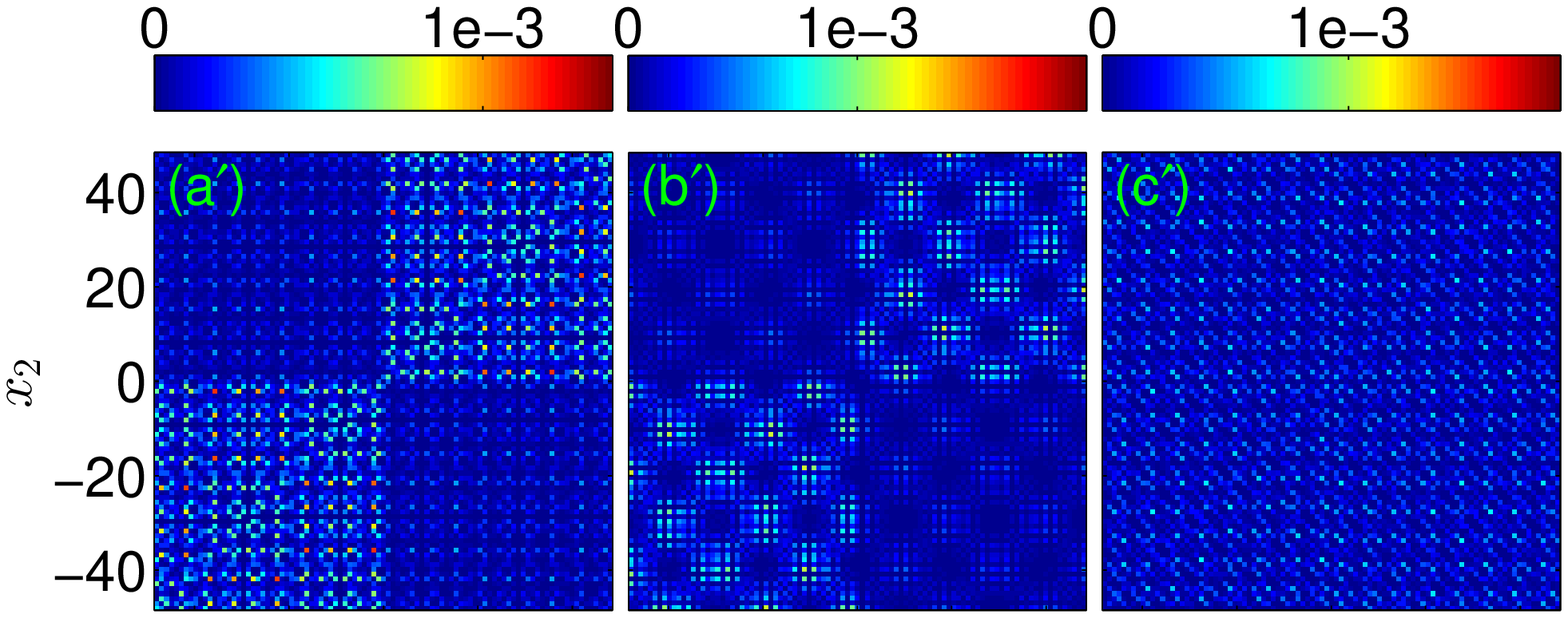}

\includegraphics[width=0.45\textwidth]{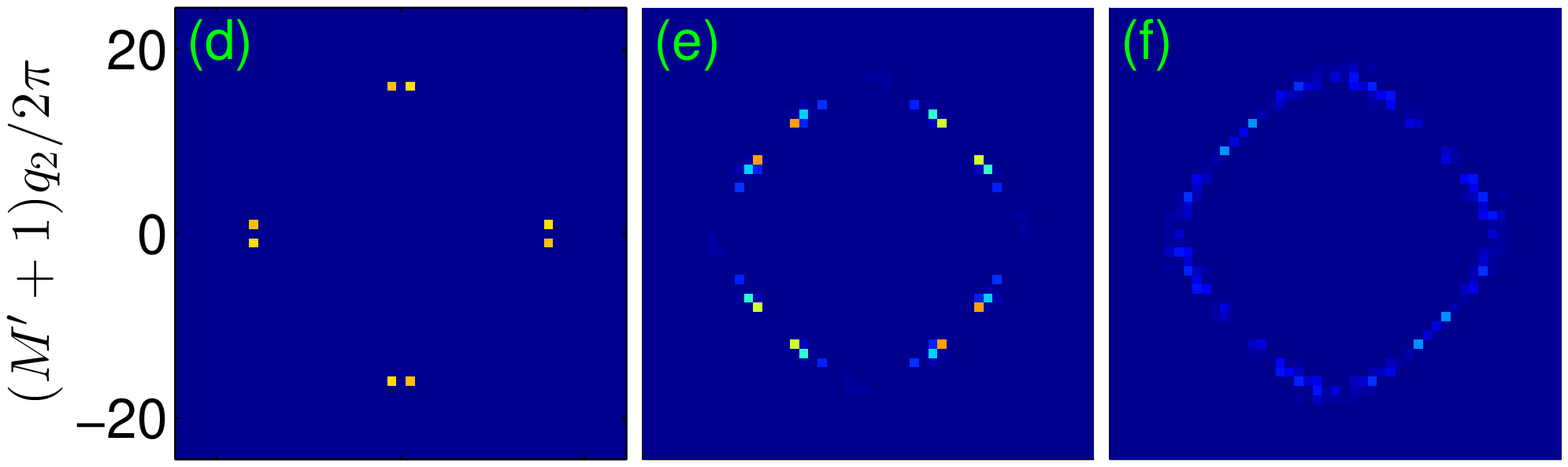}
\includegraphics[width=0.45\textwidth]{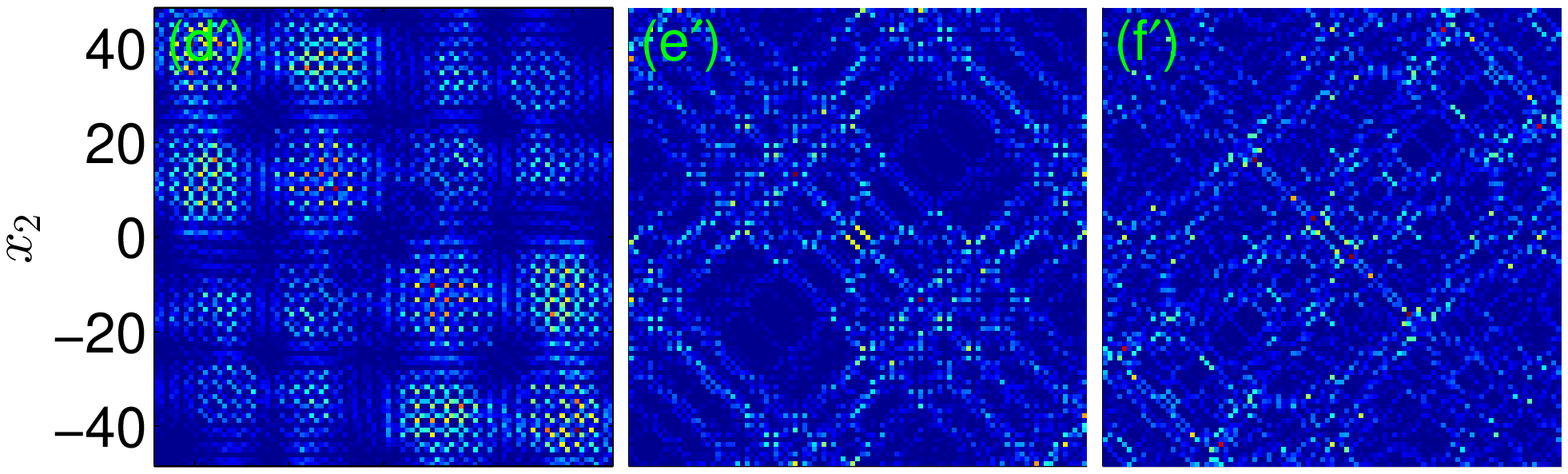}

\includegraphics[width=0.45\textwidth]{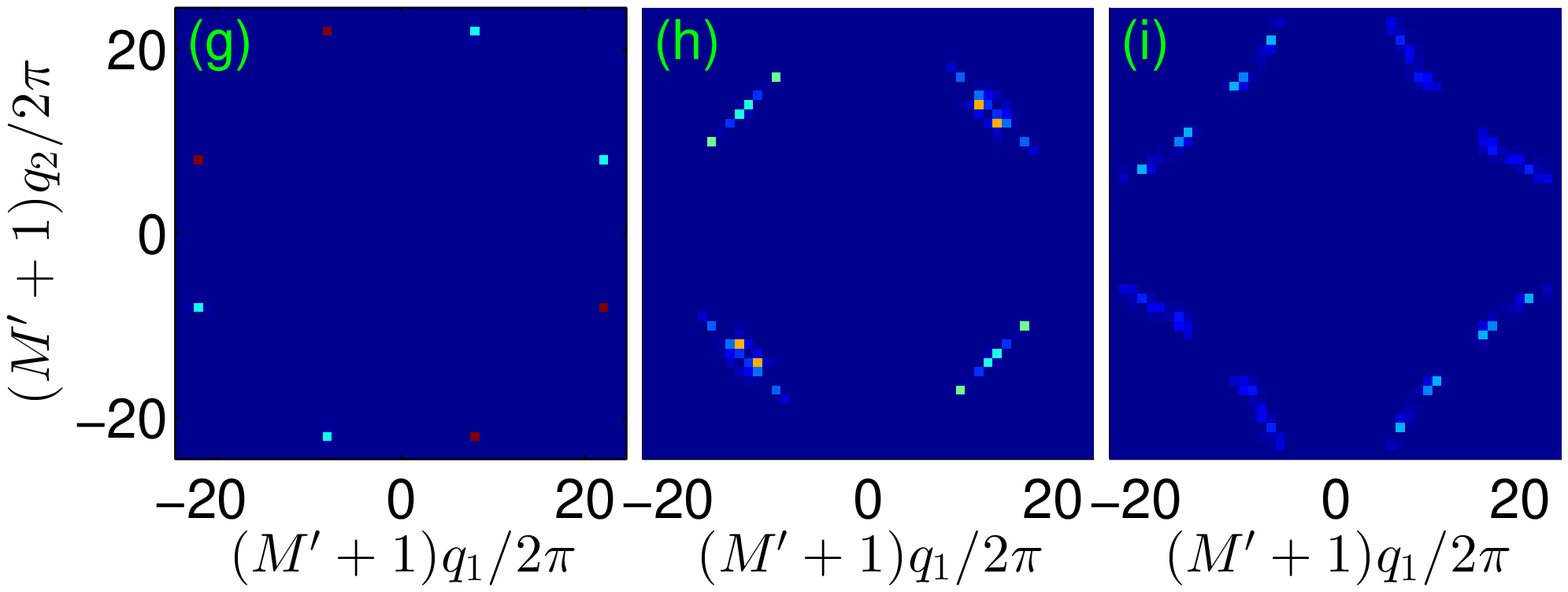}
\includegraphics[width=0.45\textwidth]{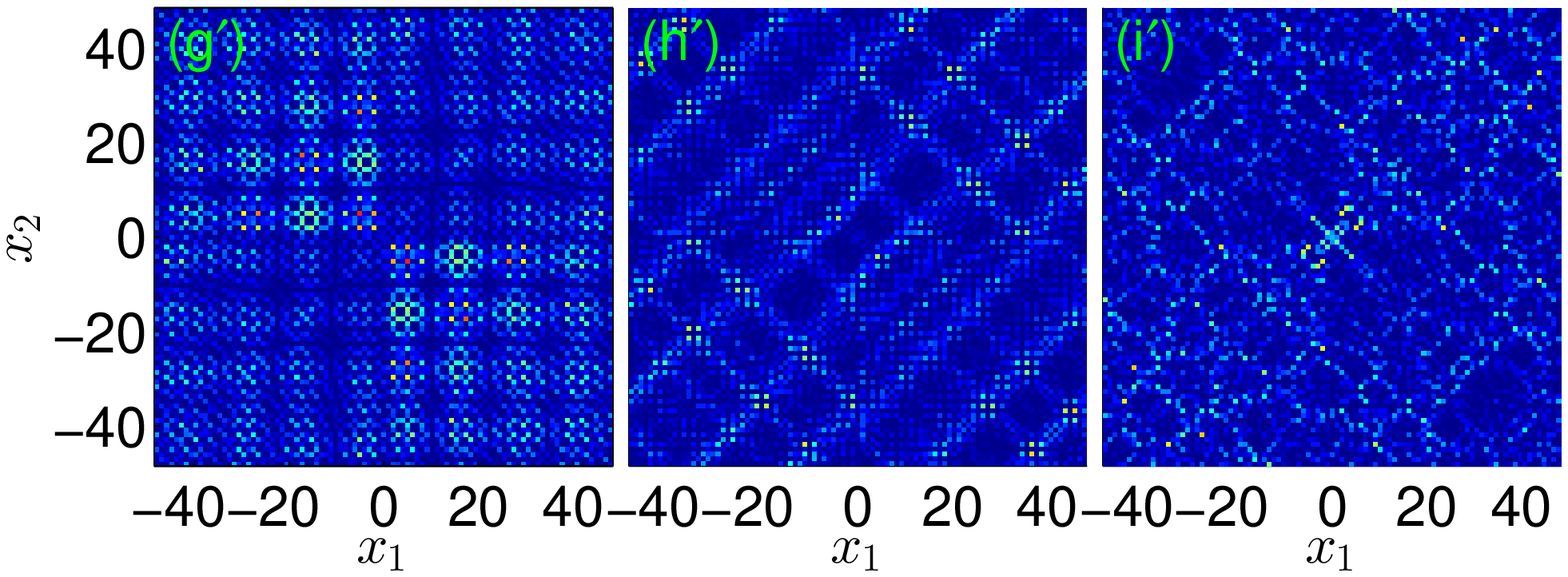}
\caption{\label{moms}(Color online) The wave function in region $[102]$ (see Fig.~\ref{lattice}) in momentum space [left three columns, labeled as (a)-(i)], and the full wave function in real space [right three columns, labeled as (a$^\prime$)-(i$^\prime$) in correspondence]. In the left three columns, $|F(q_1, q_2)|^2$ is shown and in the right three columns $|\psi(x_1, x_2)|^2$. In the first row, the three states have odd parity and the boundary condition is open in (a) and (b), while periodic in (c). In the second and third rows, the states have even parity, and the boundary condition is open and periodic, respectively. In all panels, $(V,U,M')=(-1,-1,47)$. From (a) to (i), the energy $E$ of the corresponding state is $0.7020$, $0.7994$, $0.8600$, $-0.9580$, $-1.0134$, $-0.8033$, $0.9193$, $0.7093$, and $0.6205$, respectively. }
\end{figure*}

\begin{figure*}[thb]
\includegraphics[width=0.45\textwidth]{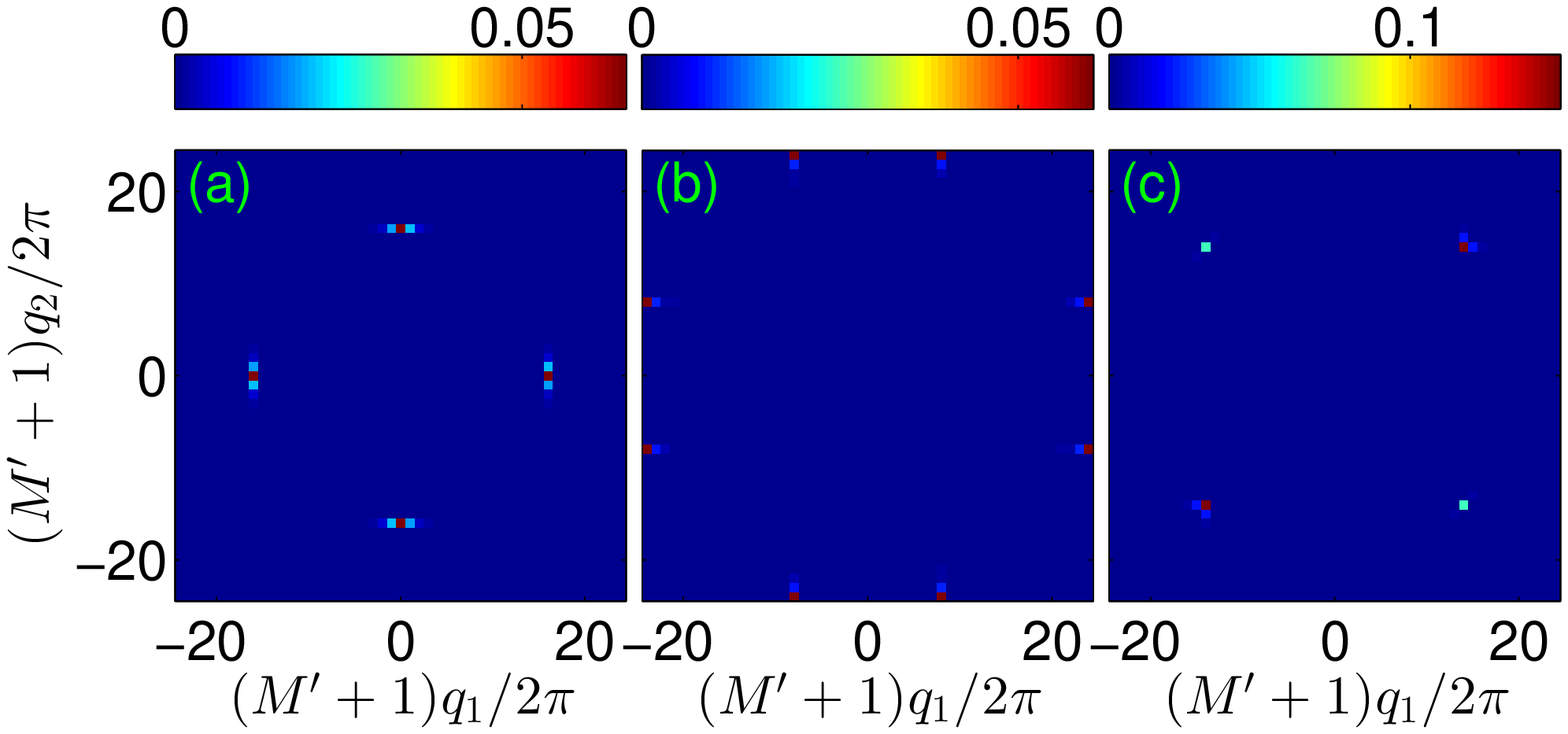}
\includegraphics[width=0.45\textwidth]{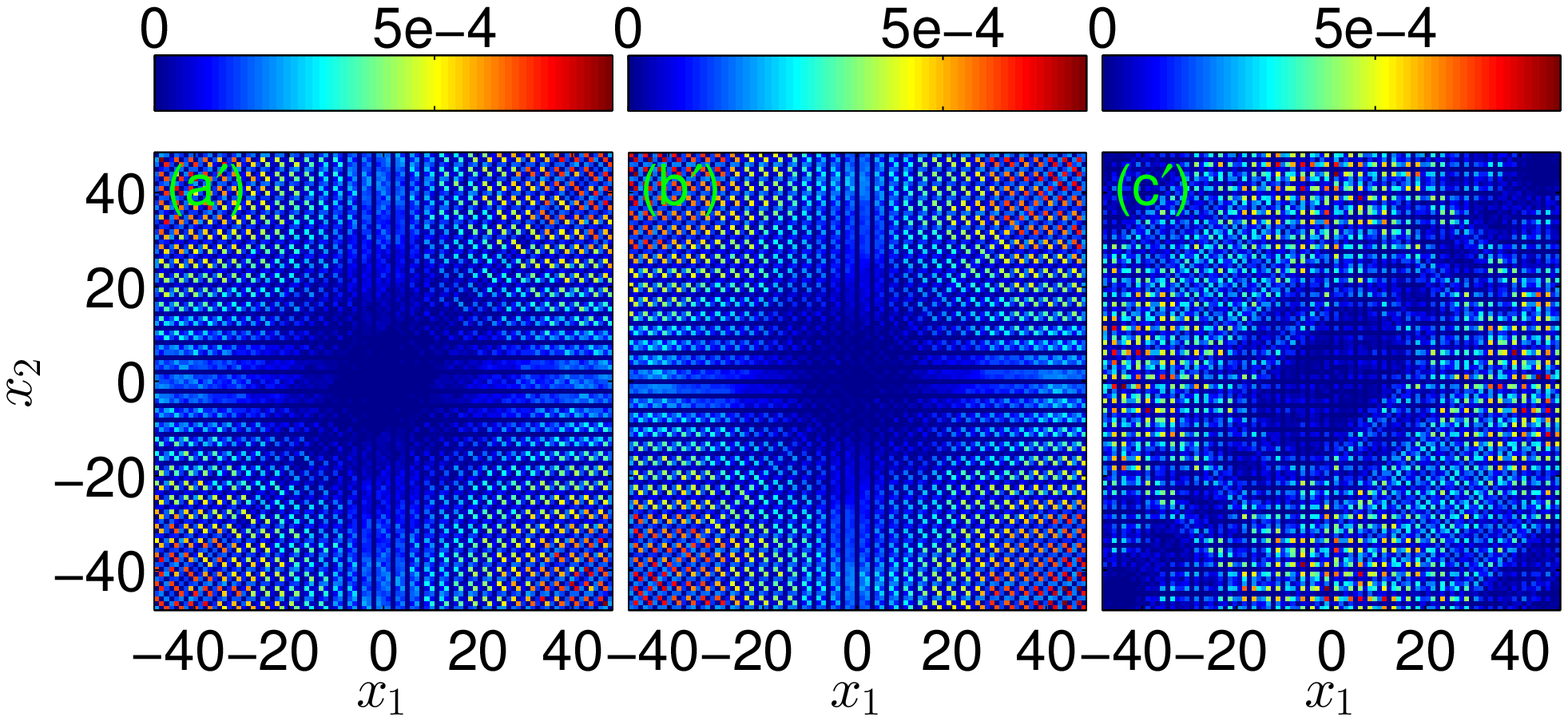}
\caption{\label{strangemoms}(Color online) More examples of weakly diffractive even-parity states. The parameters are the same as in Fig.~\ref{moms} and the boundary condition is periodic. Like in Fig.~\ref{moms}, the left three panels show the momentum distribution of the wave functions in region [102], while the right three panels show the real space density distribution of the wave functions in the whole space. The energy  of the states in (a)-(c) is $-0.9988$, $0.9989$, $1.1012$, respectively.}
\end{figure*}

The ansatz wave function (\ref{ansatz}) in each region is a superposition of eight plane waves with symmetrically related wave vectors. To check whether an eigenstate indeed has this property or not, a natural idea is to do Fourier analysis and have a look of the momentum distribution of the wave function directly. Fortunately, this idea can be readily carried out in region $[102]$, whose rectangular shape means that we have a set of plane waves as a complete orthonormal basis. 
We thus perform the (discrete) Fourier transform of the wave function in region $[102]$ as follows:
\beq
F(q_1, q_2)= \frac{1}{M'+1}\sum_{x_1=-M'}^0\sum_{x_2=0}^{M'} \psi(x_1,x_2) e^{-i q_1 x_1 -i q_2 x_2}
\label{Fq1q2}
\eeq
where $q_i = 2\pi n_i /(M'+1)$ ($i=1,2$), with $n_i =0,1,\cdots, M'$. Of course, $q_i$ are defined only up to an integer multiple of $2\pi $.
In the following, we will take them in the interval $[-\pi ,\pi)$. 

\subsubsection{Weakly diffractive even-parity states}{\label{weakdiff}}
The simple Fourier analysis reveals interesting facts. 

For the odd-parity states, the eight plane waves are beautifully demonstrated in Figs.~\ref{moms}(a)-(c). In these three panels, we see eight sharp peaks symmetrically distributed \cite{sharp}. From one peak, by inversion about the origin, reflection about the $q_1$, $q_2$ axes as well as the $q_1=\pm q_2$ axes, we can get all the eight peaks. 
The Bethe form of the odd-parity states is thus vividly demonstrated. 
We have also had a look of the wave functions in real space as in Figs.~\ref{moms}(a$^\prime$)-(c$^\prime$). They are apparently very regular.  

For the even-parity states, something unexpected emerges. Because the Yang-Baxter equation is violated for the even-parity states \cite{pra2013}, diffraction is inevitable and the eight peaks are expected to disappear or at least be strongly smeared. However, for some states, they remain and are remarkably sharp. Two such states are shown in Figs.~\ref{moms}(d) and \ref{moms}(g), with open and periodic boundary condition, respectively. We see that the eight peaks look even sharper than their counterparts in Figs.~\ref{moms}(a)-(b), and they are symmetrically distributed as in the odd-parity case. Actually, in Figs.~\ref{moms}(d) and \ref{moms}(g), the eight red squares contribute to 0.6609 and 0.6621 of the total weight of the corresponding wave function (in region $[102]$), respectively. Therefore, though the wave function (in region $[102]$) has not exactly the form (\ref{ansatz}) (as can be verified by using the Prony algorithm above), it can be well approximated by it and is qualitatively close to it. 

We call such a state a \textit{weakly diffractive} state \cite{lamacraft}, since the diffraction-free picture survives to some extent and remains the most prominent feature of the wave function. 
Of course, not all even states are weakly diffractive. As shown in Fig.~\ref{moms}(e)-(f) and \ref{moms}(h)-(i), some have a rather broad distribution in momentum space. 
In Figs.~\ref{moms}(e) and \ref{moms}(h), the eight peaks are still prominent but the diffractive components are also apparent. In Figs.~\ref{moms}(f) and \ref{moms}(i), the eight peaks are smeared out completely and the momentum components form four continuous arks. Of course, the arks are just the energy isolines of a free particle in a two dimensional square lattice because all scattering is elastic. 

In Fig.~\ref{strangemoms}, we show another three weakly diffractive states. The common feature of them is that there are not eight, but four peaks in the momentum distribution. One should note that in panel (b), two squares on opposite sides of the Brillouin zone should be identified. 

The weakly diffractive states remind us of the quantum scar states appearing in some quantized classically chaotic billiards \cite{heller}. There, although most eigenstates manifest apparent chaotic features, some do exhibit very regular behavior. 

\section{The generalized Yang-Baxter equation in the even sector}\label{YBE}

The weakly diffractive states in Figs.~\ref{moms}(d) and \ref{moms}(g) can be understood by examining the generalized YBE \eqref{YBE2} quantitatively. As mentioned in Sec.~\ref{framework}, the generalized YBE \eqref{YBE2} is violated in the even sector. We find explicitly for the difference
\begin{equation}
\S_D=\S_{10}\S_{12}'\S_{20}\S_{12}-\S_{12}\S_{20}\S_{12}'\S_{10}
\label{sdiff}
\end{equation}
the expression ($s_j=\sin k_j$)
\begin{equation}
\S_D\!=\!\frac{-2iUVs_1s_2 \hat{\s}_y}{(s_1-i\Vh)(s_2-i\Vh)(s_1-s_2+i\frac{U}{2})(s_1+s_2-i\frac{U}{2})}, \nonumber
\end{equation}
where $\hat{\s}_y$ is the usual Pauli matrix.

We see that $\S_D$ vanishes if $U$ or $V$ vanishes. In the opposite limit of $\max\{|U|,|V|\}= \infty $, ${\S_D}$ vanishes again.  Under these circumstances, the model is fully integrable. For finite $(U,V)$, $\S_D$ vanishes also if $\sin k_1=0$ or $\sin k_2=0$. This case is forbidden in the ansatz \eqref{ansatz} because the scattering at the impurity would lead to a vanishing wave function as $e^{ikx}=e^{-ikx}$ for $k=0,\pi$.
Therefore, there are no admissible even-parity Bethe states for one of the $k$'s being zero or $\pi$.
Nevertheless, there may be even-parity states composed of a set of plane waves with one component having
very small $\sin k$ --- the corresponding amplitudes would not become strongly attenuated during scattering
because the YBE is only weakly violated. These amplitudes could constitute therefore the dominant component of the wave function and lead to the weakly diffractive states observed in
Fig.~\ref{moms}. Indeed, the momentum distribution of 
the states in panels (d) and (g) of Fig.~\ref{moms} has peaks at values of $k_1$ (or $k_2$) close to $0$ or $\pm \pi$, respectively.

The states in Figs.~\ref{strangemoms}(a) and \ref{strangemoms}(b) can be understood as the extreme limits of the states in Figs.~\ref{moms}(d) and \ref{moms}(g), respectively. 

However, we note that the weakly diffractive state in Fig.~\ref{strangemoms}(c) cannot be understood in this way. The state apparently has $k_1 \simeq k_2$, but $\hat{S}_D$ does not vanish at $k_1=k_2$. 
The momenta $k_j$ can never coincide in the Bethe ansatz for similar reasons as $k_j=0,\pi$ are forbidden. States having approximately this  property are not ruled out in the even sector and they do appear. Moreover,  
they are not confined to special values of $U,V$ and therefore there must be a general reason for their existence. At present, this poses an open problem. 

Finally, let us note that the argument breaks down if one tries to use the standard formalism as was done in Ref.~\cite{pra2013}. There the amplitudes were defined locally and the YBE has the 
trilinear form
\beq
\S_{12}\S_{10}\S_{20}=\S_{20}\S_{10}\S_{12}.
\label{YBEold}
\eeq  
The kernel of the corresponding difference matrix $\S_D'$ is the odd subspace
\cite{pra2013}, so the integrability condition is satisfied for odd states also in this formalism. 
However, if one projects $\S_D'$ onto the even subspace, it never becomes zero, not even small, for $UV\neq 0$. 
Within this framework, all weakly diffractive states appear mysterious, as the YBE is always strongly violated.  

\section{Conclusions}\label{conclusions}
We have studied a one-dimensional Bose-Hubbard model with a single defect
in the two-particle sector on finite lattices both analytically and numerically. The system, though simple, shows a rich variety of properties and peculiar features, some of them appearing counter-intuitive. The impurity potential should render the model nonintegrable because incoming waves are partly transmitted and partly reflected at the central site. Nevertheless, half of the eigenstates have the Bethe ansatz form, i.e. they are
characterized by only two wave numbers $k_1,k_2$.  
Similar to systems with boundary potentials,
only the absolute values of $k_1,k_2$ are good quantum numbers. 
To analyze the eigenstates properly, the Bethe ansatz for spinless particles with scalar scattering phases has to be generalized by the introduction of non-commutative scattering matrices and non-local amplitudes. This formalism is adapted to incorporate the $\Zz_2$-symmetry (parity), which is probably the source of the partial integrability of our model. Only odd-parity states conform to the Bethe ansatz, whereas the even-parity states show diffraction and avoided crossings, a hallmark of non-integrable systems.

We have derived the generalized Bethe ansatz equations and proven that their solutions span the complete odd subspace for periodic boundary conditions. Moreover we have studied the model numerically for periodic and open boundary conditions
by using a novel technique to check for the presence of non-diffractive states after exact diagonalization. This technique is numerically exact 
and very general. It can be used in many situations if one suspects that a model may be (fully or partially) diffractionless
but cannot be treated analytically. 
We have used the method to confirm full integrability in a variant of our 
model introduced by Longhi and Della Valle  \cite{longhi} --- the infinite lattice is replaced by a semi-infinite one and the impurity sits at the end. 
Putting open boundary conditions at the other end, we have studied this system with the checking algorithm and find that all states have the Bethe form.
This comes not unexpected, as many boundary potentials do not destroy integrability \cite{sklyanin}.

Furthermore, we have analyzed the numerically obtained eigenstates in both sectors via Fourier transform to extract the momentum distribution. This gives another (approximate) criterion to test whether a given state is close to the Bethe ansatz.
We confirm in this way the integrable nature of the odd sector. In the even sector, another unexpected phenomenon appears: Some of the states show a momentum distribution with eight strong peaks, very similar to what one would expect for a state in Bethe form. The peaks are located near $k=0,\pi$, where the generalized Yang-Baxter relation is satisfied also in the even sector. Now the values $k=0,\pi$ are forbidden within the Bethe ansatz formalism and we are confronted with a novel object: eigenstates which have ``almost'' the Bethe form and can therefore be called ``weakly'' diffractive. Interestingly, the standard approach employing local amplitudes does not show this connection between a weakly violated Yang-Baxter relation and weakly diffractive states, because the former is strongly violated for arbitrary momenta. 

A natural direction of further research would be the question whether systems with an integrable sector due to the presence of a non-local symmetry (in our case space inversion with respect to the impurity site) exist also for more than two particles.

\appendix*
\section{Analytic Proof of Completeness}\label{appA}
We shall prove the full integrability of the sector with negative parity 
for odd lattice size $M$
and periodic boundary conditions by showing that the Bethe ansatz equations \eqref{BAE1} have $(M^2-1)/4$ distinct solutions. We are guided in this effort by the numerically obtained types of possible eigenstates in Sec.~\ref{3bands}.

\begin{figure}[t!]
\centering
\vspace*{0.5cm}
\includegraphics[width=0.95\columnwidth]{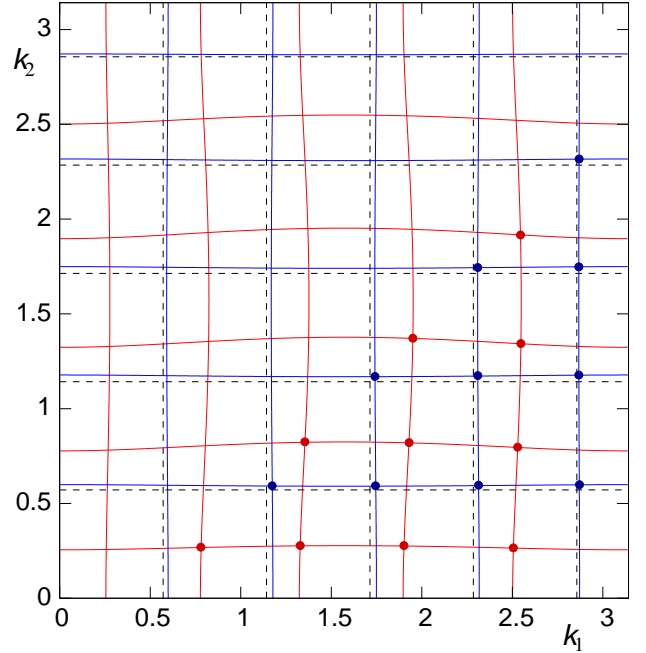}
\vspace*{-0.3cm}
\caption{(Color online)
Graphical solution of \eqref{BAE2-1}, \eqref{BAE2-2} for $U=3$, $V=-2$ and $M=11$. The blue (red) lines correspond to $\v_+$ ($\v_-$). Dashed lines denote
the values $k_j=2m_j\pi/M$ obtained for the non-interacting model.
}
\label{fig-band1}
\end{figure}

We begin by assuming both $k_1$ and $k_2$ real, corresponding to the first band.
Without loss of generality, we set $0\le k_2 <k_1$ and suppress the index $+/-$.
Taking the logarithm of \eqref{BAE1}, one obtains,
\begin{align}
k_1&=\frac{2\pi}{M}m_1 -\frac{\arg(\lo(k_1,k_2))}{M}, \label{BAE2-1}\\
k_2&=\frac{2\pi}{M}m_2 -\frac{\arg(\lz(k_1,k_2))}{M}.\label{BAE2-2}
\end{align}
The integers $m_j$ are the quantum numbers of the solution. They range over the set $0,1,\ldots M'$. The maximal number of solutions is obtained if each pair $0\le m_2\le m_1\le M'$ would correspond to an admissible solution of \eqref{BAE2-1}, \eqref{BAE2-2}. Consider the function
$\kt_1(m_1,k_2)$ obtained as solution of \eqref{BAE2-1} for given $m_1$ and and arbitrary $k_2\in[0,\pi]$. Because $-\pi\le\arg(\lo(k_1,k_2))\le\pi$,
we have $[(2m_1-1)\pi/M\le\kt_1(m_1,k_2)\le(2m_1+1)\pi/M$ for $m_1>1$ and
   $0\le\kt_1(0,k_2)\le\pi/M$ for $m_1=0$. Drawn in the $k_1/k_2$-plane, the
$\kt_1(m_1,k_2)$ form at most $M'+1$ different lines located in the stripes
$\st(m_1)=\{k_1\in I_{m_1}, k_2\in[0,\pi]\}$ with $I_{m_1}=[(2m_1-1)\pi/M,(2m_1+1)\pi/M]$ for $m_1\ge 1$ 
and  $I_0=[0,\pi/M]$ for $m_1=0$. In the same way,
the $M'+1$ graphs of the functions $\kt_2(m_2,k_1)$ occupy an orthogonal set of stripes in the  $k_1/k_2$-plane. Their crossing points correspond to a solution of the coupled equations \eqref{BAE2-1} and \eqref{BAE2-2}. 
Because of the symmetry $\lo(k_2,k_1)=\lz(k_1,k_2)$, it is clear that for 
$m_1=m_2$ the crossing point is located at $k_1=k_2$. Now it is easy to see that the wavefunction \eqref{ansatz} vanishes for $k_1=k_2$ because the scattering phase $\a(k,k)=-1$, a feature shared by all models solvable by Bethe ansatz. Therefore the set $\{(m_1,m_2)\}$ with $m_1=m_2$ has to be excluded from the list of possible quantum numbers. 
The variation of $\arg(\lo(k_1,k_2))/M$ with $k_1$ for fixed $k_2$ is of order
$M^{-2}$ in each stripe $\st(m_1)$, therefore there is exactly one solution
of \eqref{BAE2-1} for all $k_2$ if $\arg(\lo(k_1,k_2))$ is continuous as function of $k_1$ in $I_{m_1}$ and large $M$. If $\lo(k_1,k_2)$ passes through $-1$, its argument
changes discontinuously from $\pm\pi$ to $\mp\pi$. If such a jump is located at $k_1^0\in I_{m_1}$, a simple geometrical argument shows that it depends on the sign of $\partial\arg(\lo(k_1,k_2))/\partial k_1$ (which can be assumed to be constant throughout the interval
$I_{m_1}$ for sufficiently large $M$) whether a solution exist or not. If $-\partial\arg(\lo(k_1,k_2))/\partial k_1>0$, there is no solution to \eqref{BAE2-1} ($\arg(\lo(k_1))$ jumps from $-\pi$ to $\pi$)
and if  $-\partial\arg(\lo(k_1,k_2))/\partial k_1<0$ we have two solutions
($\arg(\lo(k_1))$ jumps from $\pi$ to $-\pi$).  
Now $\arg(\lo(k_1))$ is symmetric with respect to $\pi/2$, depending on $k_1$ only through $\sin k_1$, 
therefore 
$\partial\arg(\lo(k_1,k_2))/\partial k_1|_{k_1^0}=-\partial\arg(\lo(k_1,k_2))/\partial k_1|_{\pi-k_1^0}$. 
If the Interval $I_{m_1}$ contains no solutions because
of a discontinuity at $k_1^0$, the interval $I_{m_1'}$ containing $\pi-k_1^0$ will contain two solutions. This argument is valid for all $1\le m_1<M'$, which means that the sum of the number of solutions in intervals $I_1\ldots I_{(M-3)/2}$ equals $(M-3)/2$. The intervals $I_0$ and $I_{M'}$ require a separate treatment because $\lo_\pm(0,k_1)=\lo_\pm(\pi,k_2)=\mp\sign(V)$ and 
$-\arg(\lo)$ is either zero or $\pm\pi$ at these points.    
If it is zero, the interval $I_0$ contains no solution and $I_{M'}$ contains one solution. If $-\arg(\lo(0,k_2))=\pi$, $I_0$ contains one solution
and $I_{M'}$ none. For $-\arg(\lo(0,k_2))=-\pi$ the situation is reversed. We conclude that the total number of solutions to \eqref{BAE2-1} for fixed $k_2$ equals $M'$. The solutions to \eqref{BAE2-1}, \eqref{BAE2-2}  
pertaining to $\v_+$, resp. $\v_-$ are parametrized by the quantum numbers $(m_1,m_2)$ with $m_2<m_1$ and $m_1,m_2$ may take 
$M'$ different values. This leads altogether to $2(M-1)(M-3)/8$
solutions in the first band.
Fig.~\ref{fig-band1} shows an example for $M=11$.

Compared with the total number 
$(M^2-1)/4$ of states in the sector with negative parity, $M-1$ states are missing. These are states with complex momenta and are located in the second and third band.  

\begin{figure}[t!] 
\centering
\vspace*{0.5cm}
\includegraphics[width=0.95\columnwidth]{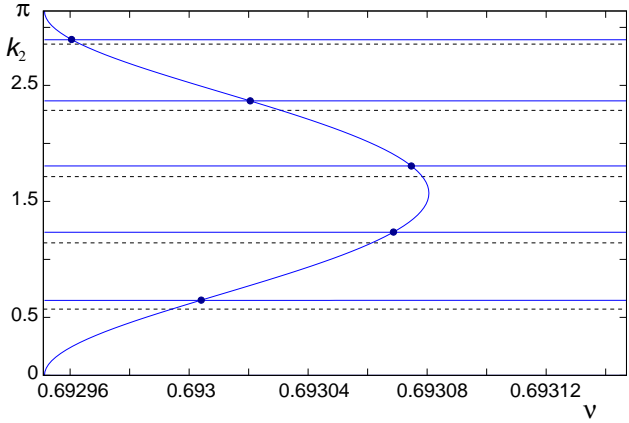}
\vspace*{-0.3cm}
\caption{(Color online)
Graphical solution of \eqref{BAE1} for the states in the second band
and parameters $U=3$, $V=-1.5$ and $M=11$  . 
The limiting value of $\ka$ for $M\ra\infty$ is $\asinh(0.75)\approx0.6931$. 
All solutions belong to $\v_+$.
Dashed lines denote
the values $k_2=2m\pi/M$ obtained for the non-interacting model.
}
\label{fig-band2}
\end{figure}

\begin{figure}[t!] 
\centering
\vspace*{0.5cm}
\includegraphics[width=0.95\columnwidth]{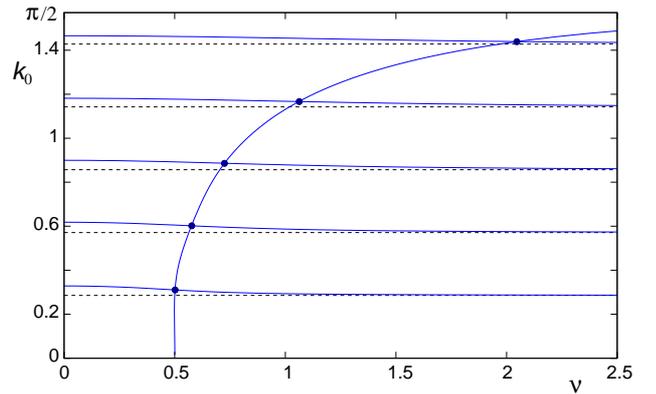}
\vspace*{-0.3cm}
\caption{(Color online)
Graphical solution of \eqref{BAE1} for the states in the third (molecule) band
and parameters $U=-2$, $V=1$ and $M=11$. 
In this case no bound state exists in the infinite system and the third band
contains $M'=5$ states belonging to $\v_+$. 
Dashed lines denote
the values $k_0=m\pi/M$.
}
\label{fig-band3}
\end{figure}

From \eqref{lam12} one sees that $\lz_\pm(k_1,k_2)$ may stay unimodular if
$\sin k_1$ is purely imaginary, as then $a_2,b,c_2,d_2$ remain real and unimodularity follows from \eqref{identity}. States of this form constitute the second band.
This entails either $k_1=i\ka$ or $k_1=\pi+i\ka$ with $\ka$ real.
In the first case, $\sin k_1=i\sinh \ka$,
\eqref{BAE1} for $k_1$ reads then $\lo_\pm=e^{\ka M}$, i.e.  $\lo_\pm$ becomes exponentially large in $M$ for $\ka>0$. This means $c_1+id_1\sim 0$ for large $M$,
or $\sinh\ka\sim -V/2$, which entails $V<0$ as expected for a bound state
at the impurity site corresponding to the lower edge of the free band.
For $V>0$, a solution exists if  
$\sin k_1=-i\sinh\ka$, the second case above.  
To proceed, we calculate the numerator of $\lo_\pm(\ka,k_2)$
for $\ka\sim\asinh(-V/2)$, $V<0$ (first case),
\beq
a_1\pm ib\sim i\frac{|V|}{2}\left(\frac{U^2-V^2}{4}-s_2^2\pm
\left|s_2^2-\frac{U^2-V^2}{4}\right|\right).
\label{numlam}
\eeq
It follows that for $s_2^2>(U^2-V^2)/4$ the numerator of $\lo_+(\ka,k_2)$
vanishes as well as the denominator at $\ka=\asinh(|V|/2)$ excluding $\v_+$.
Similarly, for  $s_2^2<(U^2-V^2)/4$, $\v_-$ is excluded: For all values of the parameters $U,V$ and $k_2$ only one of the vectors $\v_\pm$ may contribute to a state in the second band. The same conclusion holds for $\ka\sim\asinh(V/2)$, $V>0$. 
Let $\tilde{\ka}(k_2)$ denote the solution of 
$e^{\ka M}=\lo(\ka,k_2)$ for the appropriate $\v_\pm$ as function of
$k_2\in [0,\pi]$. For large $M$, $\tilde{\ka}(k_2)$ will lie in a bounded interval $I_\ka$ close to the limiting value $\asinh(|V|/2)$.
The equation for the $k_2$ reads as before
\beq
k_2= \frac{2\pi}{M}m -\frac{\arg(\lz(\ka,k_2))}{M}.
\label{BA2b}
\eeq
 We obtain again $M'+1$ stripes $\{k_2\in I_m,\ka\in I_\ka\}$. The same argument as above (which was independent from the value of $k_2$, whose role is now played by $\ka$) shows that 
$M'$ solutions of \eqref{BA2b} exist as function of $\ka$.
Because $\tilde{\ka}(k_2)$ is bounded for $0\le k_2\le \pi$, there exist
always $M'$ crossing points as admissible solutions of \eqref{BAE1}. 
Fig.~\ref{fig-band2} presents an example of the case $U^2-V^2>4s_2^2$, in which all solutions belong to $\v_+$.
The third band is composed of states with complex conjugated momenta
$k_1=k_0+i\ka$, $k_2=k_0-i\ka$. This type is 
well-known as ``string solution'' in most integrable systems.
In our case, it is not obvious that the above ansatz is consistent given the complicated form of the $\l^{(j)}_\pm$ in \eqref{lam12}.
From \eqref{BAE1} follows that ${\lz_\pm}^*=1/\lo_\pm$, which can be checked using $s_2=s_1^*$ and the identity \eqref{identity}. The ansatz is therefore consistent. Taking $\ka>0$, the denominator of $\lo_\pm(k_0,\ka)$ must tend to zero
for $M\ra\infty$, leading now to the relation
$\cos k_0\sinh\ka\sim-U/4$. This means that $k_0<\pi/2$ for $U<0$ and 
$k_0>\pi/2$ for $U>0$.  
The numerator of $\lo_\pm(k_0,\ka)$ reads then
\beq
a_1\pm ib\sim\left(\sin^2k_0\cosh^2\ka+\frac{U^2}{16}\right)[-iU\pm i|U|].
\label{numlam2}
\eeq
For $U<0$, the numerator of $\lo_-$ vanishes, lifting the divergence, and for $U>0$, the numerator of
$\lo_+$ is zero. Again only one of the $\v_\pm$ yields  solutions of \eqref{BAE1}, similar to the second band. The function $\tilde{\ka}(k_0)$  is now defined only in one of the intervals $I_<=[0,\pi/2]$ ($U<0$) and
$I_>=[\pi/2,\pi]$ ($U>0$). To determine $k_0$ we consider
\beq
\lo(k_0,\ka)\lz(k_0,\ka)=e^{-2ik_0M}
\label{BA3}
\eeq
from which we obtain
\beq
k_0=\frac{\pi}{M}m -\frac{\arg(\lo\lz)}{2M}.
\label{BA3b}
\eeq
Now the $M'+1$ stripes have width $\pi/M$ and are located in $I_<$ or 
$I_>$. One may again infer that there are $M'$ admissible
solutions $\tilde{k}_0(\ka)$ and therefore at most $M'$ crossing points with the function $\tilde{\ka}(k_0)$. However, because the latter is not bounded
in $k_0$, diverging in the vicinity of $k_0=\pi/2$, it is not clear whether
the third band contains indeed $M'$ solutions or only $M'-1$. Fig.~\ref{fig-band3} shows an example where the former is the case - for parameters at which no bound state
exists in the infinite system.
From the analysis in \cite{pra2013} we know that at most  one bound state  
may appear in the infinite system. This cannot change in the finite case (i.e. there can be no more bound states for finite $M$ than for $M\ra\infty$). This means that at most one state from the third band can become a bound state in certain parameter ranges, which depend for finite $M$ not only on $U,V$ but also on $M$
as seen in Fig.~\ref{region}.

It follows that the Bethe ansatz equations have at least 
$(M^2-1)/4-1$ solutions in the three bands considered. The missing state
is a bound state where both $\sin k_1$ and $\sin k_2$ are purely imaginary.
This state exists only for negative $U,V$ with $|U|<2|V|$. A more detailed analysis could reveal the exact conditions (including the dependence on $M$) under which a molecule state from the third band transforms into a bound state.
Here we have confined ourselves to the proof that all states in the sector with negative parity are solutions 
of \eqref{BAE1}.

\section*{Acknowledgments}

We are grateful to Eoin Quinn and Shi-Dong Jiang for helpful comments. This work was supported in part by the Deutsche Forschungsgemeinschaft through TRR 80.


\begin{thebibliography}{99}
\bibitem{sharp}
The eight peaks are not as sharp as located on a single square, simply because the wave vectors $k_{1,2}$ are generally not integer multiples of $2\pi/(M'+1)$, as can be seen in Fig.~\ref{k1k2}(a).
\end{thebibliography}
\end{document}